\documentclass[aps,prb,reprint,groupedaddress,amsmath,showpacs]{revtex4-1}

 \usepackage{graphicx}
 \usepackage{amssymb}
\usepackage[
    bookmarks,
    bookmarksopen=true,
    colorlinks=true,
    linkcolor=red, 
    anchorcolor=black,
    citecolor=blue, 
    filecolor=cyan, 
    menucolor=red, 
    urlcolor=blue,
    plainpages=false, 
    pdfpagelabels, 
    hypertexnames=false, 
    linktocpage 
]{hyperref}

\newcommand{\abs}[1]{\left| #1 \right|} 
\newcommand{\avg}[1]{\left< #1 \right>} 
 
\let\baraccent=\= 
\renewcommand{\=}[1]{\stackrel{#1}{=}} 
\newcount\colveccount
\newcommand*\colvec[1]{
        \global\colveccount#1
        \begin{pmatrix}
        \colvecnext
}
\def\colvecnext#1{
        #1
        \global\advance\colveccount-1
        \ifnum\colveccount>0
                \\
                \expandafter\colvecnext
        \else
                \end{pmatrix}
        \fi
}

\begin{document}


\title{Mach-Zehnder interferometry with periodic voltage pulses}


\author{Patrick P. Hofer}
\email[]{patrick.hofer@unige.ch}
\author{Christian Flindt}
\affiliation{D\'epartement de Physique Th\'eorique, Universit\'e de Gen\`eve, CH-1211 Gen\`eve, Switzerland}


\date{\today}

\begin{abstract}
We investigate theoretically a Mach-Zehnder interferometer driven by a time-dependent voltage. Motivated by recent experiments, we focus on a train of Lorentzian voltage pulses which we compare to a sinusoidal and a constant voltage. We discuss the visibilities of Aharonov-Bohm oscillations in the current and in the noise. For the current, we find a strikingly different behavior in the driven as compared to the static case for voltage pulses containing multiple charges. For pulses containing fractional charges, we find a universality at path-length differences equal to multiples of the spacing between the voltage pulses. These observations can be explained by the electronic energy distribution of the driven contact. In the noise oscillations, we find additional features which are characteristic to time-dependent transport. Finite electronic temperatures are found to have a qualitatively different influence on the current and the noise.
\end{abstract}


\maketitle



\section{Introduction}

\begin{figure}[t!]
\centering
\includegraphics[width=.8\columnwidth]{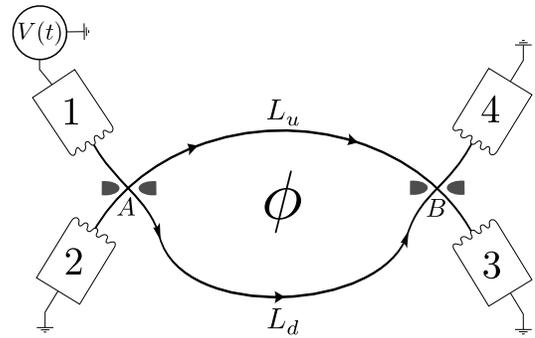}
\caption{Mach-Zehnder interferometer driven by a time dependent voltage $V(t)$. The interferometer consists of two paths of lengths $L_u$ and $L_d$ which enclose a magnetic flux $\phi$. Two quantum point contacts $A$ and $B$ allow the single particle states to be delocalized over both paths leading to interference contributions in current and noise measured at the outputs.}
  \label{fig:mzi}
\end{figure}

The electronic Mach-Zehnder interferometer\cite{ji:2003} (MZI, see Fig.~\ref{fig:mzi}) is a powerful tool to probe interference effects of individual electrons in mesoscopic conductors. The injected electrons are delocalized over two paths, leading to oscillations in the outgoing current as a function of the enclosed magnetic flux.  Working in the quantum Hall regime, where transport occurs along chiral edge states,\cite{buttiker:1988} each electron traverses the interferometer only once. The current oscillations in the MZI have been measured with a visibility above fifty percent.\cite{ji:2003,neder:2007nat,neder:2007natp,neder:2007,roulleau:2008prl100,litvin:2010,helzel:2012}
In addition, the MZI has been employed in a variety of experiments with the objective to control dephasing and decoherence. These include measuring\cite{roulleau:2008prl100} and improving\cite{huynh:2012} the coherence length in quantum Hall systems as well as tuning the decoherence using a voltage probe.\cite{roulleau:2009} Further experiments have measured the transmission phase of a quantum dot\cite{litvin:2010} and controlled the dephasing using an additional detector channel.\cite{neder:2007,neder:2007natp,roulleau:2008prl101} Finally, by inverting the role of detector and system, signatures of the noise\cite{neder:2007natp,roulleau:2008prl101,neder:2007iop} and the full counting statistics\cite{helzel:2012,levkivskyi:2009} of a quantum point contact (QPC) have been observed using a MZI.

On top of these experimental advances, the realization of driven single-electron emitters has recently paved the way for giga-hertz quantum electronics. Single electrons can now be emitted into a coherent conductor using either a mesoscopic capacitor or designed voltage pulses. The mesoscopic capacitor\cite{feve:2007,moskalets:2008} emits a sequence of electrons and holes in response to an external ac modulation. By applying a train of Lorentzian-shaped voltage pulses to an ohmic contact, noiseless excitations can be created on top of the Fermi sea in a mesoscopic conductor.\cite{dubois:2013}
These clean few-electron excitations were proposed by Levitov and co-workers\cite{levitov:1996,keeling:2006} and have recently been named levitons following their experimental realization.\cite{dubois:2013}

The combination of these achievements will surely lead to rich physics in the giga-hertz regime. Theoretical proposals for future experiments include measurements of the Glauber correlation function,\cite{haack:2011,haack:2013} the observation of interference fringes in the current and in the electronic energy distribution,\cite{ferraro:2013prb} and the modification of interferences using an additional single-electron source.\cite{juergens:2011,rossello:2014} Further theoretical studies investigate a MZI driven by a quantum pump\cite{chung:2007} and the charge transmitted through a MZI biased with Lorentzian voltage pulses.\cite{gaury:2014}

In this work, we consider a MZI driven by a time-dependent voltage. Motivated by the experimental realization of levitons, we focus on Lorentzian-shaped voltage pulses and compare our results to a sinusoidal and a constant voltage.\cite{chung:2005} The current at the outputs depends on the shape of the applied voltages if we use a MZI with a path-length difference. Such an asymmetric MZI constitutes an energy-dependent scatterer. In addition to the current visibility, we investigate the noise produced by the periodically driven MZI and find a contribution with no dc counterpart as predicted in Ref.~\onlinecite{battista:2014}. For the visibility of the current as well as the noise oscillations, we identify a lobe structure which contains information about the energy distribution of the electrons in a driven contact.

We note that a similar lobe structure was measured as a function of the applied dc voltage.\cite{neder:2006,roulleau:2007,neder:2007natp,litvin:2008,huynh:2012,helzel:2012} 
These observations are attributed to interactions within the same\cite{kovrizhin:2009} or with a neighboring edge channel.\cite{sukhorukov:2007,levkivskyi:2008} Since the interactions depend on the number of electrons in the MZI but not on their energy distribution, we employ a noninteracting scattering approach\cite{moskalets:book} at filling factor $\nu=1$ and focus on the effect of dephasing through a path-length difference. Loss of phase coherence in a MZI has been investigated using voltage and dephasing probes\cite{pilgram:2006,forster:2007,marquardt:2004prl,marquardt:2004prb,chung:2005} and as a result of internal potential fluctuations\cite{seelig:2001} and fluctuating environments.\cite{forster:2005,marquardt:2004prl,marquardt:2004prb,marquardt:2005} 

The rest of the paper is structured as follows. In Sec.~\ref{sec:2}, we introduce the setup and the treatment of a periodically driven contact. In Sec.~\ref{sec:3} we specify the voltage pulses and their respective energy distributions. The current produced by the driven MZI is discussed in Sec.~\ref{sec:current} and the noise in Sec.~\ref{sec:noise}. Finally, our conclusions are drawn in Sec.~\ref{sec:conclusions}.

\section{Driven Mach-Zehnder interferometer}
\label{sec:2}
Our setup is sketched in Fig.~\ref{fig:mzi}. It consists of a MZI in the quantum Hall regime biased with a periodic voltage $V(t)=V(t+\mathcal{T})$ at contact $1$. Electrons injected from contacts $1$ and $2$ are partitioned into two chiral edge channels at QPC $A$. The two edge channels enclose a magnetic flux $\phi$ and are repartitioned into contacts $3$ and $4$ at QPC $B$. The QPCs are described by the scattering matrices
\begin{equation}
\label{eq:qpc}
S_\alpha=\begin{pmatrix}
i\sqrt{\mathcal{R}_\alpha} & \sqrt{\mathcal{D}_\alpha}\\
\sqrt{\mathcal{D}_\alpha} & i\sqrt{\mathcal{R}_\alpha}
\end{pmatrix},
\end{equation}
with $\alpha=A,B$. The phases of the reflection and transmission amplitudes can be incorporated into a shift of the magnetic flux $\phi$ and are thus omitted.

Due to interference between the two paths, the current and noise measured at the outputs of the MZI oscillate as functions of $\phi$. In addition to this energy-independent phase, we consider an energy-dependent phase due to a difference in the path lengths $L_u$ and $L_d$. We assume a linear dispersion relation $E=\hbar v_Dk$ with drift velocity $v_D$ in the energy window of interest. The path-length difference is then characterized by the time 
\begin{equation}
\label{eq:pathlength}
\tau=(L_d-L_u)/v_D,
\end{equation}
which we assume to be positive without loss of generality. In addition to an overall phase-shift, the energy-dependent phase reduces the visibility of the oscillations depending on how the states of the charge carriers are distributed in energy. One of the main objectives of this work is to relate the energy distribution induced by a train of voltage pulses to the visibility of the current and noise oscillations in the MZI.

To describe the driven contact, it is convenient to split the voltage in a time-independent dc part and a time dependent ac part\cite{vanevic:2007}
\begin{equation}
\label{eq:voltagesplit}
V(t)=V_{dc}+V_{ac}(t).
\end{equation}
A dc voltage can be incorporated by a shift in the chemical potential. To treat the ac voltage
we resort to the Floquet scattering matrix approach.\cite{moskalets:2002,moskalets:book} We assume that the potential in contact $1$ is uniform and completely screened from the rest of the MZI. The voltage drop is assumed to be smooth on length-scales comparable to the Fermi wavelength in order not to induce additional scattering.\cite{pretre:1996,pedersen:1998,dubois:2013prb} The solution to the time-dependent (single-particle) Schr\"odinger equation in contact $1$ then reads
\begin{align}
\label{eq:floqwav}
&\psi_E(t)=\psi_E^0(t)e^{-i\varphi(t)},\\\label{eq:phase}
&\varphi(t)=\frac{e}{\hbar}\int\limits_{0}^{t}dt'V_{ac}(t'),
\end{align}
where $\psi_E^0(t)$ is a solution of the Schr\"odinger equation for $V=V_{dc}$. We note that the lower limit in the last integral can be shifted by a global phase shift. It is assumed that the ac voltage varies on a time scale which is sufficiently slow, such that the distribution of the electronic states is not disturbed.\cite{pretre:1996,pedersen:1998,dubois:2013prb} Although energy is no longer a good quantum number, the states $\psi_E(t)$ are then distributed according to the Fermi distribution corresponding to a dc bias. The distributions in the different contacts read
\begin{subequations}
\label{eq:fermidist}
\begin{align}
&f_1(E)=f(E-eV_{dc}),\\
&f_{i\neq1}(E)=f(E)=\frac{1}{e^{(E-\mu)/k_BT}+1},
\end{align}
\end{subequations}
where the index $i=1,2,3,4$ denotes the contact.
Here $T$ is the electronic temperature and $\mu$ the chemical potential of the grounded contacts. Note that $f_1(E)$ is the distribution of the $\psi_E(t)$ which do \textit{not} have a well defined energy due to the time-dependent phase induced by the ac voltage. Within these approximations, a contact driven by a voltage can thus be described as being dc biased with a time-dependent scattering phase $\varphi(t)$ that each particle picks up upon leaving the contact. Floquet scattering theory states that a time-dependent scattering phase scatters particles of energy $E$ into energy\cite{moskalets:book}
\begin{equation}
\label{eq:en}
E_n=E+n\hbar\Omega,
\end{equation}
with the amplitude
\begin{equation}
\label{eq:floqsource}
S_n=\int\limits_{0}^{\mathcal{T}}\frac{dt}{\mathcal{T}}e^{in\Omega t}e^{-i\varphi(t)}.
\end{equation}
Here $\mathcal{T}=2\pi/\Omega$ is the spacing between the voltage pulses
\begin{subequations}
\begin{align}
&V(t+\mathcal{T})=V(t),\\
\label{eq:phiper}
&\varphi(t+\mathcal{T})=\varphi(t).
\end{align}
\end{subequations}

With this description of the driven contact, we can write down the Floquet scattering matrix for the whole setup. For later convenience, we also write the energy of the incoming particles in the form of Eq.~\eqref{eq:en}. The amplitudes which relate incoming particles at energy $E_m$ to outgoing particles at energy $E_n$ then read
\begin{widetext}
\begin{equation}
\label{eq:floqmzi}
\begin{aligned}
\mathcal{F}_{31}(E_n,E_m)&= S_{n-m}\left[i\sqrt{\mathcal{R}_A\mathcal{D}_B}e^{i(\phi_u+t_u E_n/\hbar)}+i\sqrt{\mathcal{D}_A\mathcal{R}_B}e^{i(\phi_d+t_d E_n/\hbar)}\right]=S_{n-m}S_{31}(E_n),\\
\mathcal{F}_{32}(E_n,E_m)&=\delta_{n,m}\left[\sqrt{\mathcal{D}_A\mathcal{D}_B}e^{i(\phi_u+t_u E_n/\hbar)}-\sqrt{\mathcal{R}_A\mathcal{R}_B}e^{i(\phi_d+t_d E_n/\hbar)}\right]=\delta_{n,m}S_{32}(E_n),\\
\mathcal{F}_{41}(E_n,E_m)&= S_{n-m}\left[-\sqrt{\mathcal{R}_A\mathcal{R}_B}e^{i(\phi_u+t_u E_n/\hbar)}+\sqrt{\mathcal{D}_A\mathcal{D}_B}e^{i(\phi_d+t_d E_n/\hbar)}\right]=S_{n-m}S_{41}(E_n),\\
\mathcal{F}_{42}(E_n,E_m)&=\delta_{n,m}\left[i\sqrt{\mathcal{D}_A\mathcal{R}_B}e^{i(\phi_u+t_u E_n/\hbar)}+i\sqrt{\mathcal{R}_A\mathcal{D}_B}e^{i(\phi_d+t_d E_n/\hbar)}\right]=\delta_{n,m}S_{42}(E_n),
\end{aligned}
\end{equation}
\end{widetext}
where only the differences $t_{d}-t_u=\tau$ and $\phi_d-\phi_u=\phi$ matter in the following. The Floquet amplitudes $S_n$ are given in Eq.~\eqref{eq:floqsource} and the quantities $S_{ij}$ are the scattering matrices of the static MZI.\cite{chung:2005}

\section{Periodic voltage pulses}
\label{sec:3}
Motivated by recent experiments,\cite{dubois:2013,jullien:2014} we focus on voltage pulses of Lorentzian shape and compare them to sinusoidal pulses and a dc voltage. Explicitly, the voltages read
\begin{subequations}
\label{eq:voltages}
\begin{align}
\nonumber
&eV^L(t)=\sum\limits_{j=-\infty}^{\infty}\frac{2q\hbar\Gamma}{\left(t-j\mathcal{T}\right)^2+\Gamma^2}\\&\hspace{1cm}=q\hbar\Omega\frac{\sinh(\Omega\Gamma)}{\cosh(\Omega\Gamma)-\cos(\Omega t)},\\
&eV^S(t)=q\hbar\Omega\left[1+\cos\left(\Omega t\right)\right],\\\label{eq:voltagesc}
&eV^{dc}=q\hbar\Omega.
\end{align}
\end{subequations}
Here $\Gamma$ parametrizes the width of the Lorentzian pulses and $q$ is the average charge that is emitted by the contact with each pulse
\begin{equation}
\label{eq:charge}
\int\limits_{0}^{\mathcal{T}}I(t)dt=\int\limits_{0}^{\mathcal{T}}\frac{e^2}{h}V(t)dt=eq.
\end{equation}
For all voltages, the dc part is thus given by Eq.~\eqref{eq:voltagesc}. For Lorentzian pulses, the limit $\Gamma\rightarrow\infty$ corresponds to the dc case.
A case of special interest is provided by Lorentzian voltage pulses with integer charge ($q=\pm1,\pm2,...$). In this case, the current is carried by clean $q$-particle excitations,\cite{levitov:1996,keeling:2006,dubois:2013prb} termed levitons, with no accompanying electron-hole pairs. For positive (negative) $q$, the levitons are made up only of electrons above (holes below) the chemical potential.

For the voltages given in Eqs.~\eqref{eq:voltages}, the Floquet amplitudes induced by the ac voltage read (for Lorentzian pulses only the case $q=1$ is given)
\begin{subequations}
\label{eq:floqsourceexp}
\begin{align}
&S_n^L=\begin{cases}
2e^{-(n+1)\Omega\Gamma}\sinh(\Omega\Gamma) &\mbox{ if } n>-1,\\
-e^{-\Omega\Gamma} &\mbox{ if } n=-1,\\
0 &\mbox{ if } n<-1,
\end{cases}\\
&S_n^S=J_{n}(q)\\
&S_n^{dc}=\delta_{n,0},
\end{align}
\end{subequations}
where we used the Bessel function of the first kind
\begin{equation}
\label{eq:bessel}
J_n(x)=\frac{1}{2\pi}\int\limits_{-\pi}^{\pi}e^{i[nt-x\sin(t)]}dt.
\end{equation}
For a dc bias, the result simply states that there is no ac voltage that could change the energy of the electrons. For levitons of integer charge $q>0$, $S_n^L$ is only finite for $n\geq-q$. For this reason the resulting excitations are purely electronic. Similarly, for negative integer pulses, $S_n^L$ is only finite for $n\leq \abs{q}$ and the excitations are made up of holes alone.\cite{keeling:2006,dubois:2013prb}

\begin{figure*}[t!]
\centering
\includegraphics[width=1\textwidth]{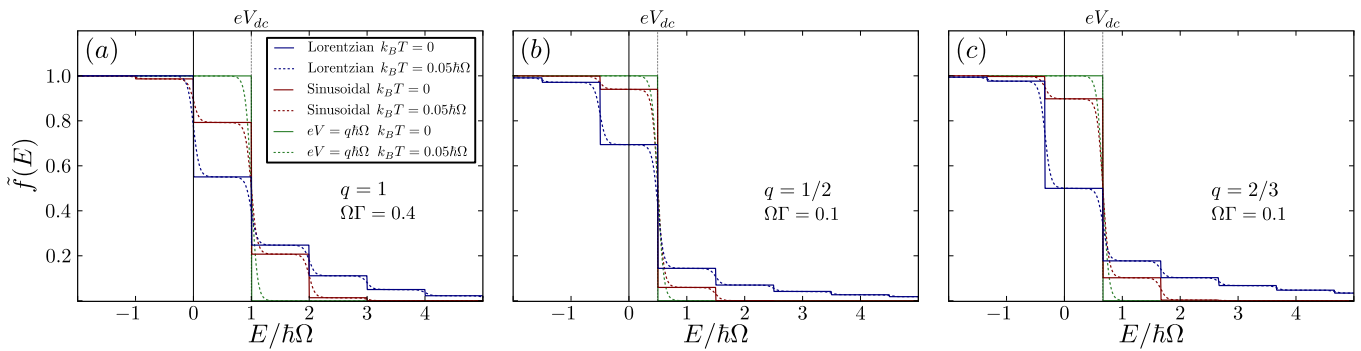}
\caption{Energy distribution of a driven contact. The filling is plotted for a contact driven by Lorentzian (blue) and sinusoidal (red) voltage pulses as well as a constant bias (green). The three panels $(a)$-$(c)$ show different pulse-charges $q$ for zero (solid) and finite (dotted) temperatures. For voltages applied at $\Omega=2\pi\cdot6$\,GHz, the dotted curves correspond to $T\approx14 $\,mK. The filling shows plateaus of width $\hbar\Omega$ for all charges $q$. For fractional charges, the steps are shifted by $eV_{dc}=q\hbar\Omega$ with respect to $\mu=0$. The widths of the pulses $\Gamma$ are chosen to best visualize the step-like behavior of the filling.}
  \label{fig:filling}
\end{figure*}

For a better understanding of the driven contact, we discuss the energy distribution of the particles leaving the contact. To this end, we introduce the second quantized operators describing particles before and after the scattering phase $\hat{a}(E)$, $\hat{b}(E)$. These are related as
\begin{equation}
\label{eq:secqsource}
\hat{b}(E)=\sum_{n=-\infty}^{\infty}S_n\hat{a}(E_{-n}),
\end{equation}
where $E_n$ is given in Eq.~\eqref{eq:en}. Using
\begin{equation}
\avg{\hat{a}^\dagger(E)\hat{a}(E_n)}=\delta_{0,n}f_1(E_n),
\end{equation}where $f_1(E)$ is given in Eqs.~\eqref{eq:fermidist}, we find
\begin{equation}
\label{eq:distsource}
\avg{\hat{b}^\dagger(E)\hat{b}(E_n)}=\sum\limits_{m=-\infty}^{\infty}S^*_mS_{m+n}f_1(E_{-m}).
\end{equation}
Particles at energies that do not differ by an integer multiple of $\hbar\Omega$ are not correlated. We will call the diagonal part ($n=0$) of the last expression the filling
\begin{equation}
\label{eq:filling}
\tilde{f}(E)=\avg{\hat{b}^\dagger(E)\hat{b}(E)}=\sum\limits_{m=-\infty}^{\infty}\abs{S_m}^2f_1(E_{-m}),
\end{equation}
and the off-diagonal terms ($n\neq0$) the coherences of the driven contact. The filling for voltage pulses of Lorentzian and sinusoidal shapes is plotted in Fig.~\ref{fig:filling} for different charges and temperatures. A finite filling above $\mu$ indicates electrons and a filling less than unity below $\mu$ indicates holes emitted by contact $1$. In contrast to the static case, a periodic voltage thus induces transport mediated by electrons and holes.\cite{battista:2014}

At zero temperature, the last equation shows that the filling in general will be constant over the energy range $\hbar\Omega$. A finite temperature smears these steps. This is confirmed by Fig.~\ref{fig:filling} which shows the filling for different charges. For fractional charges, the steps are shifted with respect to the chemical potential $\mu=0$ and there is a step at $eV_{dc}=q\hbar\Omega$. As shown below, this shift manifests itself in the visibilities of the current and noise oscillations. For the sinusoidal case, the filling exhibits the symmetry $\tilde{f}(eV_{dc}+E)=1-\tilde{f}(eV_{dc}-E)$ which is a direct consequence of the symmetry $J_{-n}(x)=(-1)^nJ_n(x)$ of the Bessel functions.

As we will see below, the dc current of the MZI is determined by the filling alone whereas the zero-frequency noise is influenced by the coherences if scattering is energy-dependent. This is the case for a MZI with a finite path-length difference [$\tau\neq0$, cf.~Eq.~\eqref{eq:pathlength}]. Both the filling and the coherences of levitons have recently been measured.\cite{dubois:2013,jullien:2014}

Finally, it is important to note that a periodically driven contact is different from a system with an equivalent filling that could be provided by an array of beam splitters and contacts at different voltages. While the latter system would describe a noisy channel, the zero frequency noise of a driven contact alone is zero, since its charge emission over the period $\mathcal{T}$ is determined solely by the dc voltage.

\begin{figure*}[t!]
\centering
\includegraphics[width=.9\textwidth]{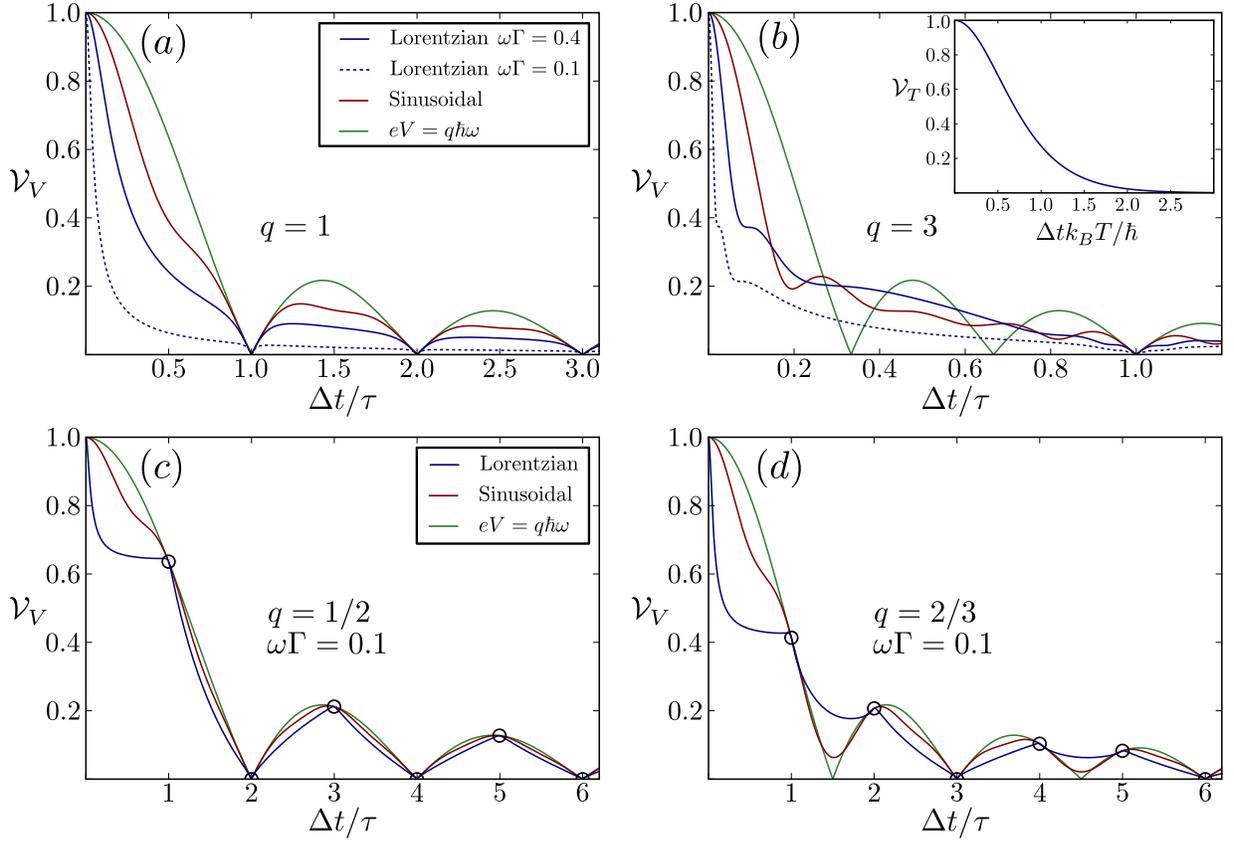}
\caption{Visibility of the current oscillations for different voltage pulses. In all cases the visibility shows an oscillatory behavior which falls of as $1/\tau$. For $q=l/p$, the visibility goes to zero if $\tau$ is an integer multiple of $p\mathcal{T}$ in the driven and of $p\mathcal{T}/l$ in the static case. A frequency $\Omega=2\pi\cdot6$\,GHz corresponds to a period $\mathcal{T}\approx0.15$\,ns which translates into a path-length difference of $\Delta L\approx1.5\,\mu$m. $(a)$ For single-charge pulses, the visibility of the driven MZI is reduced with respect to the static case but looks qualitatively similar. $(b)$ For multi-charge pulses, the visibility behaves strikingly different in the driven case. $(c)$ and $(d)$ At $\tau=j\mathcal{T}$, the visibility is independent of the pulse shape (marked by circles). All these observations are readily explained by the filling of the driven contact (see text).}
  \label{fig:visibility}
\end{figure*}

\section{Current}
\label{sec:current}
The dc current of the driven MZI in contact $3$ reads\cite{moskalets:book}
\begin{align}
\label{eq:dccurr}
\nonumber
I&=\frac{e}{h}\int\limits_{-\infty}^{\infty}dE\left[\sum\limits_{n=-\infty}^{\infty}\sum\limits_{i=1,2}\abs{\mathcal{F}_{3i}(E,E_n)}^2f_i(E_n)-f(E)\right]\\
&=\frac{e}{h}\int\limits_{-\infty}^{\infty}dE\abs{S_{31}(E)}^2\left[\tilde{f}(E)-f(E)\right].
\end{align}
Here $f_i(E)$ and $f(E)$ are given in Eqs.~\eqref{eq:fermidist} and we used Eqs.~\eqref{eq:floqmzi} and the unitarity of the scattering matrices for the second equality. The dc current is thus determined only by the filling of the driven contact and the scattering matrix of the static MZI. The current can be written as the sum of a classical part and an interference part which oscillates with $\phi$
\begin{subequations}
\label{eq:currclphi}
\begin{align}
&I=I_{cl}+I_\phi,\\
&I_{cl}=\frac{qe}{\mathcal{T}}(\mathcal{R}_A\mathcal{D}_B+\mathcal{D}_A\mathcal{R}_B),\\
\label{eq:currclphic}
&I_\phi=\frac{qe}{\mathcal{T}}2\sqrt{\mathcal{R}_A\mathcal{D}_A\mathcal{R}_B\mathcal{D}_B}\Re\left\{\avg{e^{i\Phi(E)}}\right\},
\end{align}
\end{subequations}
where $\Re\{\cdots\}$ denotes the real part. The classical part corresponds to the incoherent addition of the two paths, whereas the interference current arises from the delocalization of the single particle wavefunctions and thus depends on the phase difference of the two paths 
\begin{equation}
\label{eq:totphase}
\Phi(E)=\phi+E\tau/\hbar.
\end{equation}
The phase average in Eq.~\eqref{eq:currclphic} is defined as
\begin{equation}
\label{eq:phaseavg}
\avg{e^{i\Phi(E)}}=\frac{\int\limits_{-\infty}^{\infty}dEe^{i\Phi(E)}\left[\tilde{f}(E)-f(E)\right]}{\int\limits_{-\infty}^{\infty}dE\left[\tilde{f}(E)-f(E)\right]}.
\end{equation}
Here the denominator is proportional to the current emitted by the driven contact which is equal to $qe/\mathcal{T}$ independently of the temperature and the shape of the voltage. Note that additional dephasing can easily be incorporated in a redefinition of the phase average above (see also Ref.~\onlinecite{marquardt:2004prb}).

The visibility of the current oscillations is defined as
\begin{equation}
\label{eq:visibility}
\mathcal{V}=\frac{I_{\rm max}-I_{\rm min}}{I_{\rm max}+I_{\rm min}}=\frac{I_{\phi,\max}}{I_{cl}},
\end{equation}
where $I_{\rm max}={\rm max}_\phi I$, and similarly for $I_{\rm min}$, and we made use of $I_{\phi,\max}=-I_{\phi,\min}$ for the second equality. In our case, the phase average is proportional to $\exp{(i\phi)}$ and the interference current is maximized when the phase average becomes real and positive [cf.~Eqs.~(\ref{eq:currclphic},~\ref{eq:totphase})]. The visibility thus reads
\begin{equation}
\label{eq:visibility2}
\mathcal{V}=\frac{2\sqrt{\mathcal{R}_A\mathcal{D}_A\mathcal{R}_B\mathcal{D}_B}}{\mathcal{R}_A\mathcal{D}_B+\mathcal{D}_A\mathcal{R}_B}\abs{\avg{e^{i\Phi(E)}}}.
\end{equation}
Evaluating the phase average, we find that the temperature dependence factorizes out not only in the static,\cite{chung:2005} but also in the driven case. The visibility can therefore be written as a product of a part that depends on the QPCs, a temperature dependent part, and a part that depends on the voltage
\begin{subequations}
\label{eq:visibility3}
\begin{align}
\label{eq:visibility3a}
&\mathcal{V}=\mathcal{V}_{QPC}\cdot\mathcal{V}_T\cdot\mathcal{V}_{V},\\
&\mathcal{V}_{QPC}=\frac{2\sqrt{\mathcal{R}_A\mathcal{D}_A\mathcal{R}_B\mathcal{D}_B}}{\mathcal{R}_A\mathcal{D}_B+\mathcal{D}_A\mathcal{R}_B},\\
&\mathcal{V}_T=k_BT\frac{\pi\tau}{\hbar}{\rm csch}\left(k_BT\frac{\pi\tau}{\hbar}\right),\\\label{eq:visibility3d}
&\mathcal{V}_{V}=\abs{\sum\limits_{n=-\infty}^{\infty}\abs{S_n}^2\frac{\sin[(n+q)\Omega\tau/2]}{q\Omega\tau/2}e^{in\Omega\tau/2}}.
\end{align}
\end{subequations}
For an energy-independent MZI ($\tau=0$), the visibility is independent of temperature and voltage $\mathcal{V}_T=\mathcal{V}_V=1$ and thus does not encode any information about the filling.
The dependence of the visibility on the QPCs and the temperature is extensively discussed in Ref.~\onlinecite{chung:2005}. The visibility is maximal for a symmetric MZI ($\mathcal{D}_A=\mathcal{D}_B$), where $\mathcal{V}_{QPC}=1$. A finite temperature increases the energy spread of the involved particles leading to a monotonic decrease in the visibility. The dependence of $\mathcal{V}_T$ on $\tau$ at fixed temperature is the same as its temperature dependence for a fixed $\tau$ and is plotted in the inset of Fig.~\ref{fig:visibility}$(b)$. We now focus on $\mathcal{V}_V$ to highlight the effect of the voltage. We note that the effect of temperature on our results is negligible as long as the frequency of the applied pulses is considerably larger than the electronic temperature.

Evaluating Eq.~\eqref{eq:visibility3d} we find
\begin{subequations}
\label{eq:visibilityexpl}
\begin{align}
\label{eq:visibilityexpla}
&\mathcal{V}_V^L=\frac{\abs{\sin(\Omega\tau/2)}}{\Omega\tau/2}\frac{\sqrt{2}\sinh(\Omega\Gamma)}{\sqrt{\cosh(2\Omega\Gamma)-\cos(\Omega\tau)}},\\
&\mathcal{V}_V^S=\frac{\abs{J_0[\abs{2q\sin(\Omega\tau/2)}]e^{iq\Omega\tau}-1}}{q\Omega\tau},\\\label{eq:visibilityexplc}
&\mathcal{V}_V^{dc}=\frac{\abs{\sin(q\Omega\tau/2)}}{q\Omega\tau/2},
\end{align}
\end{subequations}
for the three different voltages, where we used the identity
\begin{equation}
\label{eq:besselid}
\sum\limits_{n=-\infty}^{\infty}J_n^2(q)e^{in\theta}=J_0[\abs{2q\sin(\theta/2)}].
\end{equation}
For Lorentzian pulses, again only the case $q=1$ is given. Taking the limit of very broad pulses $\Omega\Gamma\gg\Omega\tau, 1$, we recover the static case, Eq.~\eqref{eq:visibilityexplc} with $q=1$, from Eq.~\eqref{eq:visibilityexpla}. For well separated pulses $\Omega\Gamma,\Omega\tau\ll 1$, we recover the result of Ref.~\onlinecite{haack:2011}
\begin{equation}
\label{eq:vislev}
\left.\mathcal{V}^L_V\right|_{\Omega\rightarrow 0}=\frac{1}{\sqrt{1+\left[\tau/(2\Gamma)\right]^2}},
\end{equation}
for the injection of a single leviton into the MZI.

Figure \ref{fig:visibility} shows the visibilities for different voltages. They all show an oscillatory behavior with a decay that goes as $1/\tau$. The zeroes of the oscillations can be understood in terms of the filling (see Fig.~\ref{fig:filling}). For a pulse charge $q=l/p$, Eq.~\eqref{eq:visibility3d} shows that the visibility vanishes when $\Omega\tau/p=0\mod2\pi$. This is the case when $\tau$ is an integer multiple of $p\mathcal{T}$. At zero temperature, the filling is constant, and does not cross $\mu$, over the energy intervals [$j\hbar\Omega/p,(j+1)\hbar\Omega/p$]. Therefore, for each particle that picks up the phase $\phi+E\tau/\hbar$ we find one particle on the same plateau that picks up the phase $\phi+E\tau/\hbar\pm\Omega\tau/(2p)$. When $\tau$ is an odd multiple of $p\mathcal{T}$, these phases differ by $\pi$ and the corresponding interference effects cancel. For $\tau=jp\mathcal{T}$ with $j$ being an even number, we analogously pair up particles at $E$ and $E\pm\hbar\Omega/(2jp)$ leading to a cancellation of the interference effects. Since electrons and holes contribute with an opposite sign to the current, it is important for the above argument to pair up particles of the same kind.
Interestingly, these zeroes persist even at finite temperatures due to the factorization of Eq.~\eqref{eq:visibility3a}.

For single-charge pulses, the visibility at finite $\tau$ for a driven contact is reduced compared to the static case [see Eqs.~\eqref{eq:visibilityexpla} and \eqref{eq:visibilityexplc} and Fig.~\ref{fig:visibility}$(a)$]. We interpret this reduction in the coherence length as a result of the bunching of charge carriers within the pulses. The effect is most obvious for levitons, where $q$ electrons are localized within each pulse. Apart from this reduction, the $q=1$ case shows similar visibilities for a driven and a static contact. For $q=3$ pulses [see Fig.~\ref{fig:visibility}$(b)$], the behavior for the driven MZI becomes strikingly different from the dc case. For small $\tau$ we still observe a fast decay similar to the $q=1$ case. For larger $\tau$, however, since three charges are emitted during one period, the visibility only goes to zero at the third zero of the dc visibility. Thus, the visibility for pulsed voltages can actually be higher than the dc visibility [cf.~Eq.~\eqref{eq:visibility3d}]. Figure \ref{fig:visibility}$(b)$ furthermore shows that the visibility for Lorentzian pulses of width $\Omega\Gamma=0.1$ is strictly smaller than the visibility for $\Omega\Gamma=0.4$. Since the limit $\Omega\Gamma\rightarrow\infty$ corresponds to the dc case, this implies that the visibility has a non-monotonic dependence on the pulse width for certain regions of $\tau$.

The same features can be seen for fractional pulses. In general, for $q=l/p$, the visibility goes to zero if $\tau$ is an integer multiple of $p\mathcal{T}$ in the driven case and of $p\mathcal{T}/l$ for a constant voltage. However, for $p\neq1$, also the points $\tau=j\mathcal{T}$ for arbitrary $j$ are special. From Eq.~\eqref{eq:visibility3d} we find that the visibility is universal and only depends on the pulse charge at these universality points
\begin{equation}
\label{eq:univers}
\mathcal{V}_V(\tau=j\mathcal{T})=\frac{\abs{\sin(j\pi q)}}{j\pi q}.
\end{equation}
The universality points are marked by circles in Figs.~\ref{fig:visibility}$(c)$ and $(d)$. As $q$ becomes smaller, they become denser and the visibility becomes increasingly independent of the pulse shape. The points of universality can again be explained by the filling. Since the filling shows steps of width $\hbar\Omega$, one can again find pairs of particles which have a phase difference of $\pi$ for all steps except the one that crosses $\mu$. Since holes acquire an additional minus sign because of their charge, they add up constructively with the electrons which are $\hbar\Omega/2$ away in energy. The number of remaining particles which contribute to the interference is independent of the step height because the contributions from electrons and holes are summed up. They give rise to the same visibility as the constant voltage $eV_{dc}=q\hbar\Omega$ which leads to the universality expressed in Eq.~\eqref{eq:univers}. It is interesting to note that the interference current becomes time-independent at the universality points (not shown).

To summarize this section, we find a visibility which strongly depends on the applied voltage and differs significantly from the static case for $q\neq1$. All the features can be explained by the filling of the driven contact.

\section{Noise}
\label{sec:noise}
We now turn to the zero-frequency cross correlator between the currents in contacts $3$ and $4$
\begin{equation}
\label{eq:noise}
\mathcal{P}=\frac{1}{2}\int_{0}^{\mathcal{T}}\frac{dt}{\mathcal{T}}\int\limits_{0}^{\infty}dt'\avg{\left\{\Delta \hat{I}_3(t),\Delta \hat{I}_4(t+t')\right\}},
\end{equation}
where $\{\cdot,\cdot\}$ is the anti-commutator and $\Delta \hat{I}_i(t)=\hat{I}_i(t)-\langle\hat{I}_i(t)\rangle$. In our four-terminal setup, where there is no direct transmission between contacts $3$ and $4$, thermal equilibrium fluctuations do not contribute to the cross-correlator.\cite{buttiker:1992} The zero frequency noise arises due to partial occupation of the outgoing states. Within the scattering matrix approach, the outgoing states are mapped onto incoming states via the scattering matrix. Partial occupation of outgoing states then translates into pairs of incoming states with different fillings which feed into the same outgoing state. For each term in the noise, there are thus two incoming states involved. In our setup, this has two consequences.

First, as described in Ref.~\onlinecite{battista:2014}, the two states can either originate from two different contacts or from the same contact. The former contribution to the noise will be called transport noise, the latter interference noise (not to be confused with the $\phi$-dependent part of the noise)
\begin{equation}
\label{eq:noisetrinf}
\mathcal{P}=\mathcal{P}_{\rm tr}+\mathcal{P}_{\rm int}.
\end{equation}
When the two states originate from the same contact, they have to originate from different energies. To end up in the same outgoing state, the charge carriers need to gain or lose energy provided by the ac voltage. The interference noise thus vanishes in the dc case.

The second consequence is due to the fact that each of the two involved states can enclose the flux $\phi$. The noise thus consists of a constant part (no particles enclose the flux), a part which oscillates as a function of $\phi$ with period $2\pi$ (one particle encloses the flux) and a part which oscillates with period $\pi$ (both particles enclose the flux)\cite{chung:2005}
\begin{equation}
\label{eq:noisephi}
\mathcal{P}=\mathcal{P}_{0}+\mathcal{P}_{\phi}+\mathcal{P}_{2\phi}.
\end{equation}  
Note that the decompositions of the noise defined in Eqs.~\eqref{eq:noisetrinf} and \eqref{eq:noisephi} are independent of each other.
Before we define and discuss the measurable visibilities of the noise oscillations, we take a closer look at the decomposition of Eq.~\eqref{eq:noisetrinf}.

The transport noise can be written as
\begin{equation}
\label{eq:noisetr1}
\begin{aligned}
\mathcal{P}_{\rm tr}=\frac{e^2}{h}\int\limits_{-\infty}^{\infty}&dE\sum\limits_{n=-\infty}^{\infty}\left[f_1(E_{-n})-f(E)\right]^2\abs{S_n}^2\\&\times\Re\left\{S_{31}^*(E)S_{32}(E)S_{42}^*(E)S_{41}(E)\right\},
\end{aligned}
\end{equation}
where the Fermi functions are defined in Eq.~\eqref{eq:fermidist} and the quantities $S_{ij}(E)$ are the scattering amplitudes of the static MZI [cf.~Eqs.~\eqref{eq:floqmzi}]. We note that at zero temperature, where $f(E)=f^2(E)$, the transport noise can be expressed solely in terms of the filling $\tilde{f}(E)$, see Eq.~\eqref{eq:filling}, and the scattering matrix of the static MZI 
\begin{equation}
\label{eq:noistrt0}
\begin{aligned}
\left.\mathcal{P}_{\rm tr}\right|_{T=0}=&\frac{e^2}{h}\int\limits_{-\infty}^{\infty}dE
\left[\tilde{f}(E)-f(E)\right]^2\\&\times\Re\left\{S_{31}^*(E)S_{32}(E)S_{42}^*(E)S_{41}(E)\right\}.
\end{aligned}
\end{equation}

\begin{figure*}[t!]
\centering
\includegraphics[width=.9\textwidth]{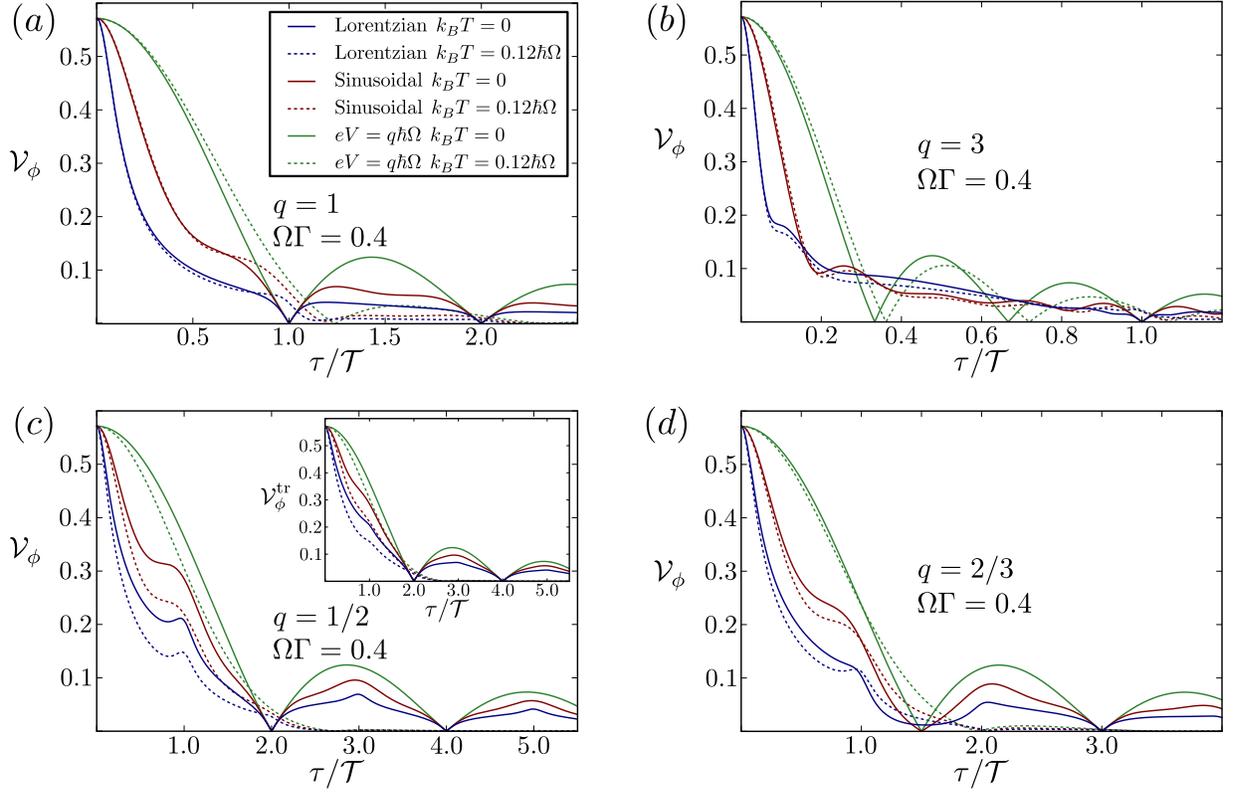}
\caption{Visibility of the noise oscillations with period $2\pi$ for different voltage pulses. In all cases the visibility shows an oscillatory behavior which falls of as $1/\tau$. The period of the oscillations is temperature dependent and only at $T=0$ the same as for the current visibility. Additional features due to the interference noise can be seen for fractional pulses, panels $(c)$ and $(d)$. They are illustrated with the help of the inset in panel $(c)$ which shows the visibility due to the transport noise alone. Here $\mathcal{D}_A=\mathcal{D}_B=0.75$. For voltages applied at $\Omega=2\pi\cdot6$\,GHz, the dotted curves correspond to $T\approx35 $\,mK.}
  \label{fig:visibilitynoise1}
\end{figure*}

\begin{figure*}[t!]
\centering
\includegraphics[width=.9\textwidth]{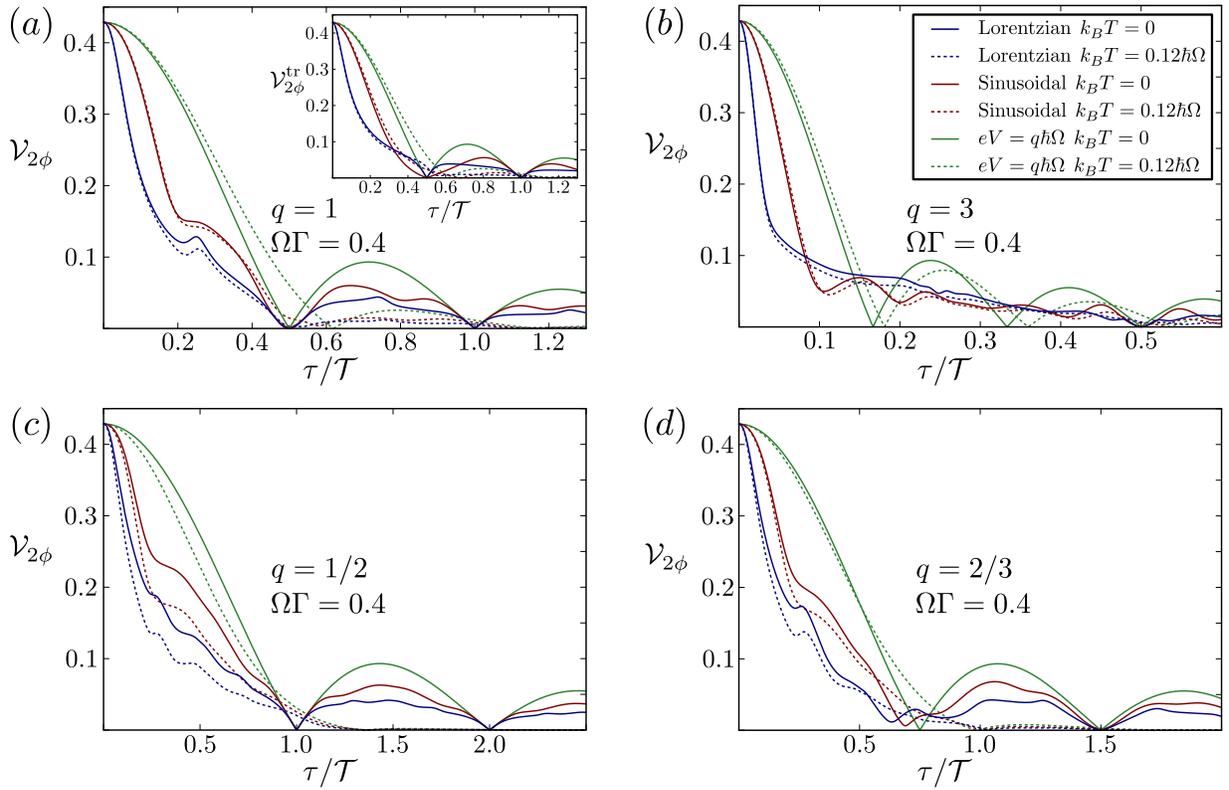}
\caption{Visibility of the noise oscillations with period $\pi$ for different voltage pulses. In all cases the visibility shows an oscillatory behavior which falls of as $1/\tau$. The period of the oscillations is half the period of $\mathcal{V}_\phi$ and is temperature dependent. Additional features due to the interference noise can be seen for all charges. These are illustrated with the help of the inset in panel $(a)$ which shows the visibility due to the transport noise alone. In contrast to the current visibility, the noise visibilities thus encode information about the coherences and cannot be explained by the filling alone. Here $\mathcal{D}_A=\mathcal{D}_B=0.75$.}
  \label{fig:visibilitynoise2}
\end{figure*}

For arbitrary temperatures, Eq.~\eqref{eq:noisetr1} can be written as
\begin{equation}
\label{eq:noisetr2}
\begin{aligned}
\mathcal{P}_{\rm tr}=-\frac{e^2}{h}&\sum\limits_{n=-\infty}^{\infty}\abs{S_n}^2\left[c_0\bar{S}_{0}^{(n+q)}\right.\\&\left.+c_{\phi}\bar{S}_{\phi}^{(n+q)}\cos[(n+q)\Omega\tau/2+\phi]\right.\\&\left.-c_{2\phi}\bar{S}_{2\phi}^{(n+q)}\cos[(n+q)\Omega\tau+2\phi]\right],
\end{aligned}
\end{equation}
where the Floquet amplitudes $S_n$ are given in Eq.~\eqref{eq:floqsource}. Here we introduced the QPC dependent coefficients
\begin{subequations}
\label{eq:noisetr3}
\begin{align}
\label{eq:noisetr3a}
&c_0=\mathcal{R}_A\mathcal{D}_A+\mathcal{R}_B\mathcal{D}_B-6\mathcal{R}_A\mathcal{D}_A\mathcal{R}_B\mathcal{D}_B,\\\label{eq:noisetr3b}
&c_{\phi}=2(\mathcal{D}_A-\mathcal{R}_A)(\mathcal{D}_B-\mathcal{R}_B)\sqrt{\mathcal{R}_A\mathcal{D}_A\mathcal{R}_B\mathcal{D}_B},\\\label{eq:noisetr3c}
&c_{2\phi}=2\mathcal{R}_A\mathcal{D}_A\mathcal{R}_B\mathcal{D}_B,
\end{align}
\end{subequations}
and the functions
\begin{subequations}
\label{eq:noisetr4}
\begin{align}
&\bar{S}_0^{(n)}=n\hbar\Omega\coth\left(\frac{n\hbar\Omega}{2k_BT}\right)-2k_BT,\\\nonumber
&\bar{S}_{j\phi}^{(n)}=\frac{2\pi k_BT}{\sinh(j\pi k_BT\tau/\hbar)}\left[\coth\left(\frac{n\hbar\Omega}{2k_BT}\right)\sin(jn\Omega\tau/2)\right.\\&\hspace{3cm}\left.-\frac{jk_BT\tau}{\hbar}\cos(jn\Omega\tau/2)\right].
\end{align}
\end{subequations}
For $S_n=\delta_{n,0}$, we recover the result of a constant voltage from Ref.~\onlinecite{chung:2005}, correcting a sign in front of the coefficient $c_{2\phi}$. We note that the part of the transport noise which oscillates with period $2\pi$ goes to zero if one of the QPCs is half-transparent [cf.~Eq.~\eqref{eq:noisetr3b}]. In this case, we can always find two processes with equal probability which only differ by an exchange of a particle originating from contact $1$ with a particle originating from contact $2$ and therefore cancel. For the part oscillating with period $\pi$, there is only one process for each pair of incoming states and it remains finite.

Already from Eq.~\eqref{eq:noisetr2}, we can anticipate that the noise visibility will not exhibit any universality. At zero temperature, the first term of Eq.~\eqref{eq:noisetr2} is proportional to the number of electrons plus the number of holes emitted by the driven contact. Their sum depends on the pulse shape unlike their difference, which only depends on the pulse charge and determines the dc current.\cite{dubois:2013prb}

The interference noise can be written as
\begin{equation}
\label{eq:noiseint1}
\begin{aligned}
\mathcal{P}_{\rm int}=&\frac{e^2}{h}\sum_{n,m,p}\int\limits_{-\infty}^{\infty}dE\frac{\left[f_1(E_{-n})-f_1(E_{-m})\right]^2}{2}\\&\times S_n^*S_mS_{m+p}^*S_{n+p}\abs{S_{31}(E)}^2\abs{S_{41}(E_p)}^2.
\end{aligned}
\end{equation}
 For the case of a constant voltage, the Floquet amplitudes imply $n=m$, which together with the Fermi functions leads to a vanishing interference noise. Similarly, for an energy-independent MZI ($\tau=0$), the sum over $p$ implies $m=n$ and the interference noise vanishes in agreement with the findings of Ref.~\onlinecite{battista:2014}.
 
 We note that even at zero temperature, the filling alone is not sufficient to describe the interference noise and the coherences become important, see Eq.~\eqref{eq:distsource},
 \begin{equation}
 \label{eq:noiseintt0}
 \begin{aligned}
 \left.\mathcal{P}_{\rm int}\right|_{T=0}=&\frac{e^2}{h}\int\limits_{-\infty}^{\infty}dE
 \abs{S_{31}(E)}^2\left[\tilde{f}(E)\abs{S_{41}(E)}^2\right.\\&-\sum\limits_{p=-\infty}^{\infty}\abs{\langle\hat{b}^\dag(E)\hat{b}(E_p)\rangle}^2\left.\abs{S_{41}(E_p)}^2\right].
 \end{aligned}
 \end{equation} 
 
At arbitrary temperatures, Eq.~\eqref{eq:noiseint1} can be cast into the more compact form
\begin{equation}
\label{eq:noiseint2}
\begin{aligned}
\mathcal{P}_{\rm int}=-\frac{e^2}{h}&c_{2\phi}\sum\limits_{n=-\infty}^{\infty}\left[\bar{S}_{0}^{(n)}\abs{\mathcal{K}_{n}}^2\right.\\&\left.+\bar{S}_{2\phi}^{(n)}\Re\left\{e^{i[2\phi+2(n+q)\Omega\tau]}\mathcal{K}_{n}\mathcal{K}_{-n}\right\}\right],
\end{aligned}
\end{equation}
having introduced
\begin{equation}
\mathcal{K}_n=\frac{1}{\sqrt{2}}\sum_{m=-\infty}^{\infty}S^*_mS_{m+n}e^{im\Omega\tau}.
\end{equation}
We note that even the $\phi$-independent part of the interference noise depends on $\tau$, since it has to vanish for $\tau=0$. We can see that the interference noise does not have a part which oscillates with period $2\pi$. Analogous to the transport noise, we always find two processes which only differ by the exchange of two particles originating from different energies. Since both particles originate from contact $1$, these processes cancel irrespectively of the QPC transmission amplitudes.

In close analogy to the current oscillations, we now define visibilities both for the part oscillating with period $2\pi$ and $\pi$
\begin{equation}
\label{eq:noisevisibilities}
\mathcal{V}_{j\phi}=\abs{\frac{\mathcal{P}_{j\phi,\max}}{\mathcal{P}_0}},
\end{equation}
where $\mathcal{P}_{j\phi,\max}=\max_\phi\mathcal{P}_{j\phi}$ [cf.~Eq.~\eqref{eq:noisephi}] and $j=1,2$. With Eqs.~(\ref{eq:noisetr2},~\ref{eq:noiseint2}) we readily find
\begin{subequations}
\label{eq:noisep}
\begin{align}
&\mathcal{P}_0=-\frac{e^2}{h}\sum\limits_{n=-\infty}^{\infty}\left[c_0\bar{S}_0^{(n+q)}\abs{S_n}^2+c_{2\phi}\bar{S}_0^{(n)}\abs{\mathcal{K}_{n}}^2\right],\\
&\mathcal{P}_{\phi,\max}=\frac{e^2}{h}\abs{\sum\limits_{n=-\infty}^{\infty}c_\phi\bar{S}_\phi^{(n+q)} e^{in\Omega\tau/2}\abs{S_n}^2},\\\nonumber
&\mathcal{P}_{2\phi,\max}=\frac{e^2}{h}\Biggl|\sum\limits_{n=-\infty}^{\infty}c_{2\phi}e^{in\Omega\tau}\left[\bar{S}_{2\phi}^{(n+q)}\abs{S_n}^2-\right.\\&\hspace{3.5cm}\left.\bar{S}_{2\phi}^{(n)}e^{i(n+q)\Omega\tau}\mathcal{K}_{n}\mathcal{K}_{-n}\right]\Biggr|.
\end{align}
\end{subequations}
For a constant voltage, $S_n=\delta_{n,0}$, we recover the results of Ref.~\onlinecite{chung:2005}
\begin{equation}
\label{eq:noisevisdc}
\mathcal{V}_{j\phi}=\abs{\frac{c_{j\phi}\bar{S}_{j\phi}^{(q)}}{c_0\bar{S}^{(q)}_0}}.
\end{equation}
The dependence of the visibilities on the QPCs is in detail discussed in Ref.~\onlinecite{chung:2005}. The ratio $\abs{c_\phi/c_0}$ is zero if one of the QPCs is either fully closed, half transparent or fully open, only going to unity for $\mathcal{D}_A,\mathcal{D}_B\ll1$. The ratio $\abs{c_{2\phi}/c_0}$ has its maximum, equal to unity, for $\mathcal{D}_A=\mathcal{D}_B=1/2$ and monotonically decreases away from that point.

The noise visibilities are plotted in Figs.~\ref{fig:visibilitynoise1} and \ref{fig:visibilitynoise2} for different pulse shapes and charges at zero and finite temperature. We find a qualitatively similar behavior to the current visibility. The noise visibilities show oscillations as a function of $\tau$ which fall of as $1/\tau$. 
Furthermore, narrower pulses as well as increasing temperatures generally lower the noise visibilities. At zero temperature, $\mathcal{V}_\phi$ goes to zero if the path-length difference $\tau$ is an integer multiple of $p\mathcal{T}$ in the driven and $p\mathcal{T}/l$ in the static case (where the charge of the current pulses is $q=l/p$). Analogous to the current visibility, this can be explained with the filling of the driven contact. Since $\mathcal{V}_{2\phi}$ oscillates at twice the frequency, it behaves similarly as a function of $2\tau$, see Fig.~\ref{fig:visibilitynoise2}.

In contrast to the current visibility, the temperature dependence of the noise visibilities does not factor out and the period of the oscillations (and thus the zeroes) depends on temperature as can be seen clearly in panels \ref{fig:visibilitynoise1}$(a)$, $(b)$ and \ref{fig:visibilitynoise2}$(a)$, $(b)$. As a consequence, a finite temperature can increase the visibilities for certain regions of $\tau$. Since the visibilities vanish at infinite temperature [cf.~Eqs.~(\ref{eq:noisetr4},~\ref{eq:noisevisibilities})], this implies a non-monotonic dependence on temperature.

Also, both $\mathcal{V}_{2\phi}$ and $\mathcal{V}_\phi$ for fractional pulses show additional features which arise due to the interference noise. In the insets of panels \ref{fig:visibilitynoise1}$(c)$ and \ref{fig:visibilitynoise2}$(a)$, we show the visibilities that arise solely from the transport noise which lack the additional features. For $\mathcal{V}_\phi$, the interference noise only influences $\mathcal{P}_0$ [cf.~Eq.~\eqref{eq:noisep}]. Since $\mathcal{K}_n$ vanishes if $\tau$ is an integer multiple of $\mathcal{T}$, $\mathcal{P}_0$ exhibits local minima at these points which leads to peaks in the visibility. These can be seen in panels \ref{fig:visibilitynoise1}$(c)$ and $(d)$, where the visibility is finite at these points due to the fractional character of the pulses. The features in $\mathcal{V}_{2\phi}$ depend on the $\mathcal{K}_n$ in a more complicated way and also arise for integer charge pulses. The noise visibilities thus clearly show signatures of the interference noise in a driven, energy-dependent MZI.

\section{Conclusions}
\label{sec:conclusions}

We have investigated the current and the noise in a MZI with a path-length difference, driven by periodic voltage pulses. Whereas the current can be expressed by the scattering matrix of the static MZI and the filling of the driven contact, the noise also depends on the coherences of the driven contact.

All the visibilities show a lobe structure which falls off as the inverse path-length difference. Quite generally, the visibilities of the current and noise oscillations decay faster with the path-length difference for narrower pulses. We interpret this as a result of a reduction in the delocalization of the electronic wavefunctions. The zeroes of the lobe structure can be explained with the filling of the driven contact and occur whenever the interference of each charge carrier is canceled by another charge carrier which picks up an additional minus sign due to the energy dependence of the MZI. Because the temperature dependence factorizes out for the current visibility, these zeroes remain robust at finite temperatures. This is not the case for the noise visibilities. The fact that the zeroes have their origin in an exact cancellation of the interference effect of different particles can lead to a non-monotonic behavior. For the current visibility, this occurs as a function of the width of the Lorentzian voltage pulses and for the noise visibilities as a function of temperature.

Additionally, the visibilities strongly depend on the charge of the voltage pulses and can behave remarkably different in the driven and the static case.
For a charge $q=l/p$, the visibilities go to zero at a path-length difference which is an integer multiple of $p\mathcal{T}$ for the driven and $p\mathcal{T}/l$ for the static case (or twice as often for $\mathcal{V}_{2\phi}$). For fractional charges, we find a universal behavior in the current visibility for a path-length difference which is an integer multiple of $\mathcal{T}$. This can be explained similarly to the visibility zeroes. At the points of universality, pairing up particles which contribute to the current oscillations with an opposite sign may leave unpaired electrons and holes. The remaining charge carriers lead to a visibility that is independent of the voltage shape. Such a universality is absent in the noise, because there electrons and holes contribute equally.

For time-dependent transport and energy-dependent scattering, the noise has an additional contribution with no dc counterpart due to processes where both involved states originate from the same (driven) contact.\cite{battista:2014} We found that this notably modulates the visibility $\mathcal{V}_\phi$ for fractional pulses and $\mathcal{V}_{2\phi}$ for all charge values.
We note that our setup can be used to implement the proposal of Ref.~\onlinecite{moskalets:2014}, where a constant scattering phase is applied to one of the (grounded) inputs of a MZI for a finite time. This scattering phase, which naturally occurs between well separated Lorentzian pulses of non-integer charge, is envisioned as a carrier of quantum information.

In summary, we have shown that the visibilities of the current and noise oscillations in a driven MZI make it possible to address different aspects of coherent time-dependent transport. We hope our work may stimulate experiments in this direction.

\begin{acknowledgments}
This work is dedicated to Markus B\"uttiker who suggested the problem. We thank F. Battista for pointing out the advantages of separating the voltage in dc and ac components and we acknowledge fruitful discussions with P. Roulleau, D. C. Glattli, G. Haack, D. Dasenbrook, P. Samuelsson, D. Ferraro, M. Moskalets and E. V. Sukhorukov. The work was funded by the Swiss NSF.
\end{acknowledgments}



\begin{thebibliography}{49}%
\makeatletter
\providecommand \@ifxundefined [1]{%
 \@ifx{#1\undefined}
}%
\providecommand \@ifnum [1]{%
 \ifnum #1\expandafter \@firstoftwo
 \else \expandafter \@secondoftwo
 \fi
}%
\providecommand \@ifx [1]{%
 \ifx #1\expandafter \@firstoftwo
 \else \expandafter \@secondoftwo
 \fi
}%
\providecommand \natexlab [1]{#1}%
\providecommand \enquote  [1]{``#1''}%
\providecommand \bibnamefont  [1]{#1}%
\providecommand \bibfnamefont [1]{#1}%
\providecommand \citenamefont [1]{#1}%
\providecommand \href@noop [0]{\@secondoftwo}%
\providecommand \href [0]{\begingroup \@sanitize@url \@href}%
\providecommand \@href[1]{\@@startlink{#1}\@@href}%
\providecommand \@@href[1]{\endgroup#1\@@endlink}%
\providecommand \@sanitize@url [0]{\catcode `\\12\catcode `\$12\catcode
  `\&12\catcode `\#12\catcode `\^12\catcode `\_12\catcode `\%12\relax}%
\providecommand \@@startlink[1]{}%
\providecommand \@@endlink[0]{}%
\providecommand \url  [0]{\begingroup\@sanitize@url \@url }%
\providecommand \@url [1]{\endgroup\@href {#1}{\urlprefix }}%
\providecommand \urlprefix  [0]{URL }%
\providecommand \Eprint [0]{\href }%
\providecommand \doibase [0]{http://dx.doi.org/}%
\providecommand \selectlanguage [0]{\@gobble}%
\providecommand \bibinfo  [0]{\@secondoftwo}%
\providecommand \bibfield  [0]{\@secondoftwo}%
\providecommand \translation [1]{[#1]}%
\providecommand \BibitemOpen [0]{}%
\providecommand \bibitemStop [0]{}%
\providecommand \bibitemNoStop [0]{.\EOS\space}%
\providecommand \EOS [0]{\spacefactor3000\relax}%
\providecommand \BibitemShut  [1]{\csname bibitem#1\endcsname}%
\let\auto@bib@innerbib\@empty
\bibitem [{\citenamefont {Ji}\ \emph {et~al.}(2003)\citenamefont {Ji},
  \citenamefont {Chung}, \citenamefont {Sprinzak}, \citenamefont {Heiblum},
  \citenamefont {Mahalu},\ and\ \citenamefont {Shtrikman}}]{ji:2003}%
  \BibitemOpen
  \bibfield  {author} {\bibinfo {author} {\bibfnamefont {Y.}~\bibnamefont
  {Ji}}, \bibinfo {author} {\bibfnamefont {Y.}~\bibnamefont {Chung}}, \bibinfo
  {author} {\bibfnamefont {D.}~\bibnamefont {Sprinzak}}, \bibinfo {author}
  {\bibfnamefont {M.}~\bibnamefont {Heiblum}}, \bibinfo {author} {\bibfnamefont
  {D.}~\bibnamefont {Mahalu}}, \ and\ \bibinfo {author} {\bibfnamefont
  {H.}~\bibnamefont {Shtrikman}},\ }\href
  {http://dx.doi.org/10.1038/nature01503} {\bibfield  {journal} {\bibinfo
  {journal} {Nature (London)}\ }\textbf {\bibinfo {volume} {422}},\ \bibinfo
  {pages} {415} (\bibinfo {year} {2003})}\BibitemShut {NoStop}%
\bibitem [{\citenamefont {B\"uttiker}(1988)}]{buttiker:1988}%
  \BibitemOpen
  \bibfield  {author} {\bibinfo {author} {\bibfnamefont {M.}~\bibnamefont
  {B\"uttiker}},\ }\href {http://link.aps.org/doi/10.1103/PhysRevB.38.9375}
  {\bibfield  {journal} {\bibinfo  {journal} {Phys. Rev. B}\ }\textbf {\bibinfo
  {volume} {38}},\ \bibinfo {pages} {9375} (\bibinfo {year}
  {1988})}\BibitemShut {NoStop}%
\bibitem [{\citenamefont {Neder}\ \emph
  {et~al.}(2007{\natexlab{a}})\citenamefont {Neder}, \citenamefont {Ofek},
  \citenamefont {Chung}, \citenamefont {Heiblum}, \citenamefont {Mahalu},\ and\
  \citenamefont {Umansky}}]{neder:2007nat}%
  \BibitemOpen
  \bibfield  {author} {\bibinfo {author} {\bibfnamefont {I.}~\bibnamefont
  {Neder}}, \bibinfo {author} {\bibfnamefont {N.}~\bibnamefont {Ofek}},
  \bibinfo {author} {\bibfnamefont {Y.}~\bibnamefont {Chung}}, \bibinfo
  {author} {\bibfnamefont {M.}~\bibnamefont {Heiblum}}, \bibinfo {author}
  {\bibfnamefont {D.}~\bibnamefont {Mahalu}}, \ and\ \bibinfo {author}
  {\bibfnamefont {V.}~\bibnamefont {Umansky}},\ }\href
  {http://dx.doi.org/10.1038/nature05955} {\bibfield  {journal} {\bibinfo
  {journal} {Nature (London)}\ }\textbf {\bibinfo {volume} {448}},\ \bibinfo
  {pages} {333} (\bibinfo {year} {2007}{\natexlab{a}})}\BibitemShut {NoStop}%
\bibitem [{\citenamefont {Neder}\ \emph
  {et~al.}(2007{\natexlab{b}})\citenamefont {Neder}, \citenamefont {Marquardt},
  \citenamefont {Heiblum}, \citenamefont {Mahalu},\ and\ \citenamefont
  {Umansky}}]{neder:2007natp}%
  \BibitemOpen
  \bibfield  {author} {\bibinfo {author} {\bibfnamefont {I.}~\bibnamefont
  {Neder}}, \bibinfo {author} {\bibfnamefont {F.}~\bibnamefont {Marquardt}},
  \bibinfo {author} {\bibfnamefont {M.}~\bibnamefont {Heiblum}}, \bibinfo
  {author} {\bibfnamefont {D.}~\bibnamefont {Mahalu}}, \ and\ \bibinfo {author}
  {\bibfnamefont {V.}~\bibnamefont {Umansky}},\ }\href
  {http://dx.doi.org/10.1038/nphys627} {\bibfield  {journal} {\bibinfo
  {journal} {Nat. Phys.}\ }\textbf {\bibinfo {volume} {3}},\ \bibinfo {pages}
  {534} (\bibinfo {year} {2007}{\natexlab{b}})}\BibitemShut {NoStop}%
\bibitem [{\citenamefont {Neder}\ \emph
  {et~al.}(2007{\natexlab{c}})\citenamefont {Neder}, \citenamefont {Heiblum},
  \citenamefont {Mahalu},\ and\ \citenamefont {Umansky}}]{neder:2007}%
  \BibitemOpen
  \bibfield  {author} {\bibinfo {author} {\bibfnamefont {I.}~\bibnamefont
  {Neder}}, \bibinfo {author} {\bibfnamefont {M.}~\bibnamefont {Heiblum}},
  \bibinfo {author} {\bibfnamefont {D.}~\bibnamefont {Mahalu}}, \ and\ \bibinfo
  {author} {\bibfnamefont {V.}~\bibnamefont {Umansky}},\ }\href {\doibase
  10.1103/PhysRevLett.98.036803} {\bibfield  {journal} {\bibinfo  {journal}
  {Phys. Rev. Lett.}\ }\textbf {\bibinfo {volume} {98}},\ \bibinfo {pages}
  {036803} (\bibinfo {year} {2007}{\natexlab{c}})}\BibitemShut {NoStop}%
\bibitem [{\citenamefont {Roulleau}\ \emph
  {et~al.}(2008{\natexlab{a}})\citenamefont {Roulleau}, \citenamefont
  {Portier}, \citenamefont {Roche}, \citenamefont {Cavanna}, \citenamefont
  {Faini}, \citenamefont {Gennser},\ and\ \citenamefont
  {Mailly}}]{roulleau:2008prl100}%
  \BibitemOpen
  \bibfield  {author} {\bibinfo {author} {\bibfnamefont {P.}~\bibnamefont
  {Roulleau}}, \bibinfo {author} {\bibfnamefont {F.}~\bibnamefont {Portier}},
  \bibinfo {author} {\bibfnamefont {P.}~\bibnamefont {Roche}}, \bibinfo
  {author} {\bibfnamefont {A.}~\bibnamefont {Cavanna}}, \bibinfo {author}
  {\bibfnamefont {G.}~\bibnamefont {Faini}}, \bibinfo {author} {\bibfnamefont
  {U.}~\bibnamefont {Gennser}}, \ and\ \bibinfo {author} {\bibfnamefont
  {D.}~\bibnamefont {Mailly}},\ }\href {\doibase
  10.1103/PhysRevLett.100.126802} {\bibfield  {journal} {\bibinfo  {journal}
  {Phys. Rev. Lett.}\ }\textbf {\bibinfo {volume} {100}},\ \bibinfo {pages}
  {126802} (\bibinfo {year} {2008}{\natexlab{a}})}\BibitemShut {NoStop}%
\bibitem [{\citenamefont {Litvin}\ \emph {et~al.}(2010)\citenamefont {Litvin},
  \citenamefont {Helzel}, \citenamefont {Tranitz}, \citenamefont
  {Wegscheider},\ and\ \citenamefont {Strunk}}]{litvin:2010}%
  \BibitemOpen
  \bibfield  {author} {\bibinfo {author} {\bibfnamefont {L.~V.}\ \bibnamefont
  {Litvin}}, \bibinfo {author} {\bibfnamefont {A.}~\bibnamefont {Helzel}},
  \bibinfo {author} {\bibfnamefont {H.-P.}\ \bibnamefont {Tranitz}}, \bibinfo
  {author} {\bibfnamefont {W.}~\bibnamefont {Wegscheider}}, \ and\ \bibinfo
  {author} {\bibfnamefont {C.}~\bibnamefont {Strunk}},\ }\href {\doibase
  10.1103/PhysRevB.81.205425} {\bibfield  {journal} {\bibinfo  {journal} {Phys.
  Rev. B}\ }\textbf {\bibinfo {volume} {81}},\ \bibinfo {pages} {205425}
  (\bibinfo {year} {2010})}\BibitemShut {NoStop}%
\bibitem [{\citenamefont {Helzel}\ \emph {et~al.}(2012)\citenamefont {Helzel},
  \citenamefont {Litvin}, \citenamefont {Levkivskyi}, \citenamefont
  {Sukhorukov}, \citenamefont {Wegscheider},\ and\ \citenamefont
  {Strunk}}]{helzel:2012}%
  \BibitemOpen
  \bibfield  {author} {\bibinfo {author} {\bibfnamefont {A.}~\bibnamefont
  {Helzel}}, \bibinfo {author} {\bibfnamefont {L.~V.}\ \bibnamefont {Litvin}},
  \bibinfo {author} {\bibfnamefont {I.~P.}\ \bibnamefont {Levkivskyi}},
  \bibinfo {author} {\bibfnamefont {E.~V.}\ \bibnamefont {Sukhorukov}},
  \bibinfo {author} {\bibfnamefont {W.}~\bibnamefont {Wegscheider}}, \ and\
  \bibinfo {author} {\bibfnamefont {C.}~\bibnamefont {Strunk}},\ }\href@noop {}
  {\enquote {\bibinfo {title} {Noise-induced phase transition in an electronic
  {M}ach-{Z}ehnder interferometer: a manifestation of non-gaussian noise},}\ }
  (\bibinfo {year} {2012}),\ \Eprint {http://arxiv.org/abs/1211.5951}
  {arXiv:1211.5951 [cond-mat]} \BibitemShut {NoStop}%
\bibitem [{\citenamefont {Huynh}\ \emph {et~al.}(2012)\citenamefont {Huynh},
  \citenamefont {Portier}, \citenamefont {le~Sueur}, \citenamefont {Faini},
  \citenamefont {Gennser}, \citenamefont {Mailly}, \citenamefont {Pierre},
  \citenamefont {Wegscheider},\ and\ \citenamefont {Roche}}]{huynh:2012}%
  \BibitemOpen
  \bibfield  {author} {\bibinfo {author} {\bibfnamefont {P.-A.}\ \bibnamefont
  {Huynh}}, \bibinfo {author} {\bibfnamefont {F.}~\bibnamefont {Portier}},
  \bibinfo {author} {\bibfnamefont {H.}~\bibnamefont {le~Sueur}}, \bibinfo
  {author} {\bibfnamefont {G.}~\bibnamefont {Faini}}, \bibinfo {author}
  {\bibfnamefont {U.}~\bibnamefont {Gennser}}, \bibinfo {author} {\bibfnamefont
  {D.}~\bibnamefont {Mailly}}, \bibinfo {author} {\bibfnamefont
  {F.}~\bibnamefont {Pierre}}, \bibinfo {author} {\bibfnamefont
  {W.}~\bibnamefont {Wegscheider}}, \ and\ \bibinfo {author} {\bibfnamefont
  {P.}~\bibnamefont {Roche}},\ }\href {\doibase 10.1103/PhysRevLett.108.256802}
  {\bibfield  {journal} {\bibinfo  {journal} {Phys. Rev. Lett.}\ }\textbf
  {\bibinfo {volume} {108}},\ \bibinfo {pages} {256802} (\bibinfo {year}
  {2012})}\BibitemShut {NoStop}%
\bibitem [{\citenamefont {Roulleau}\ \emph {et~al.}(2009)\citenamefont
  {Roulleau}, \citenamefont {Portier}, \citenamefont {Roche}, \citenamefont
  {Cavanna}, \citenamefont {Faini}, \citenamefont {Gennser},\ and\
  \citenamefont {Mailly}}]{roulleau:2009}%
  \BibitemOpen
  \bibfield  {author} {\bibinfo {author} {\bibfnamefont {P.}~\bibnamefont
  {Roulleau}}, \bibinfo {author} {\bibfnamefont {F.}~\bibnamefont {Portier}},
  \bibinfo {author} {\bibfnamefont {P.}~\bibnamefont {Roche}}, \bibinfo
  {author} {\bibfnamefont {A.}~\bibnamefont {Cavanna}}, \bibinfo {author}
  {\bibfnamefont {G.}~\bibnamefont {Faini}}, \bibinfo {author} {\bibfnamefont
  {U.}~\bibnamefont {Gennser}}, \ and\ \bibinfo {author} {\bibfnamefont
  {D.}~\bibnamefont {Mailly}},\ }\href {\doibase
  10.1103/PhysRevLett.102.236802} {\bibfield  {journal} {\bibinfo  {journal}
  {Phys. Rev. Lett.}\ }\textbf {\bibinfo {volume} {102}},\ \bibinfo {pages}
  {236802} (\bibinfo {year} {2009})}\BibitemShut {NoStop}%
\bibitem [{\citenamefont {Roulleau}\ \emph
  {et~al.}(2008{\natexlab{b}})\citenamefont {Roulleau}, \citenamefont
  {Portier}, \citenamefont {Roche}, \citenamefont {Cavanna}, \citenamefont
  {Faini}, \citenamefont {Gennser},\ and\ \citenamefont
  {Mailly}}]{roulleau:2008prl101}%
  \BibitemOpen
  \bibfield  {author} {\bibinfo {author} {\bibfnamefont {P.}~\bibnamefont
  {Roulleau}}, \bibinfo {author} {\bibfnamefont {F.}~\bibnamefont {Portier}},
  \bibinfo {author} {\bibfnamefont {P.}~\bibnamefont {Roche}}, \bibinfo
  {author} {\bibfnamefont {A.}~\bibnamefont {Cavanna}}, \bibinfo {author}
  {\bibfnamefont {G.}~\bibnamefont {Faini}}, \bibinfo {author} {\bibfnamefont
  {U.}~\bibnamefont {Gennser}}, \ and\ \bibinfo {author} {\bibfnamefont
  {D.}~\bibnamefont {Mailly}},\ }\href {\doibase
  10.1103/PhysRevLett.101.186803} {\bibfield  {journal} {\bibinfo  {journal}
  {Phys. Rev. Lett.}\ }\textbf {\bibinfo {volume} {101}},\ \bibinfo {pages}
  {186803} (\bibinfo {year} {2008}{\natexlab{b}})}\BibitemShut {NoStop}%
\bibitem [{\citenamefont {Neder}\ and\ \citenamefont
  {Marquardt}(2007)}]{neder:2007iop}%
  \BibitemOpen
  \bibfield  {author} {\bibinfo {author} {\bibfnamefont {I.}~\bibnamefont
  {Neder}}\ and\ \bibinfo {author} {\bibfnamefont {F.}~\bibnamefont
  {Marquardt}},\ }\href {http://stacks.iop.org/1367-2630/9/i=5/a=112}
  {\bibfield  {journal} {\bibinfo  {journal} {New J. Phys.}\ }\textbf {\bibinfo
  {volume} {9}},\ \bibinfo {pages} {112} (\bibinfo {year} {2007})}\BibitemShut
  {NoStop}%
\bibitem [{\citenamefont {Levkivskyi}\ and\ \citenamefont
  {Sukhorukov}(2009)}]{levkivskyi:2009}%
  \BibitemOpen
  \bibfield  {author} {\bibinfo {author} {\bibfnamefont {I.~P.}\ \bibnamefont
  {Levkivskyi}}\ and\ \bibinfo {author} {\bibfnamefont {E.~V.}\ \bibnamefont
  {Sukhorukov}},\ }\href {\doibase 10.1103/PhysRevLett.103.036801} {\bibfield
  {journal} {\bibinfo  {journal} {Phys. Rev. Lett.}\ }\textbf {\bibinfo
  {volume} {103}},\ \bibinfo {pages} {036801} (\bibinfo {year}
  {2009})}\BibitemShut {NoStop}%
\bibitem [{\citenamefont {F\`eve}\ \emph {et~al.}(2007)\citenamefont {F\`eve},
  \citenamefont {Mah\'e}, \citenamefont {Berroir}, \citenamefont {Kontos},
  \citenamefont {Pla\ifmmode~\mbox{\c{c}}\else \c{c}\fi{}ais}, \citenamefont
  {Glattli}, \citenamefont {Cavanna}, \citenamefont {Etienne},\ and\
  \citenamefont {Jin}}]{feve:2007}%
  \BibitemOpen
  \bibfield  {author} {\bibinfo {author} {\bibfnamefont {G.}~\bibnamefont
  {F\`eve}}, \bibinfo {author} {\bibfnamefont {A.}~\bibnamefont {Mah\'e}},
  \bibinfo {author} {\bibfnamefont {J.-M.}\ \bibnamefont {Berroir}}, \bibinfo
  {author} {\bibfnamefont {T.}~\bibnamefont {Kontos}}, \bibinfo {author}
  {\bibfnamefont {B.}~\bibnamefont {Pla\ifmmode~\mbox{\c{c}}\else
  \c{c}\fi{}ais}}, \bibinfo {author} {\bibfnamefont {D.~C.}\ \bibnamefont
  {Glattli}}, \bibinfo {author} {\bibfnamefont {A.}~\bibnamefont {Cavanna}},
  \bibinfo {author} {\bibfnamefont {B.}~\bibnamefont {Etienne}}, \ and\
  \bibinfo {author} {\bibfnamefont {Y.}~\bibnamefont {Jin}},\ }\href
  {http://www.sciencemag.org/content/316/5828/1169.abstract} {\bibfield
  {journal} {\bibinfo  {journal} {Science}\ }\textbf {\bibinfo {volume}
  {316}},\ \bibinfo {pages} {1169} (\bibinfo {year} {2007})}\BibitemShut
  {NoStop}%
\bibitem [{\citenamefont {Moskalets}\ \emph {et~al.}(2008)\citenamefont
  {Moskalets}, \citenamefont {Samuelsson},\ and\ \citenamefont
  {B\"uttiker}}]{moskalets:2008}%
  \BibitemOpen
  \bibfield  {author} {\bibinfo {author} {\bibfnamefont {M.}~\bibnamefont
  {Moskalets}}, \bibinfo {author} {\bibfnamefont {P.}~\bibnamefont
  {Samuelsson}}, \ and\ \bibinfo {author} {\bibfnamefont {M.}~\bibnamefont
  {B\"uttiker}},\ }\href
  {http://link.aps.org/doi/10.1103/PhysRevLett.100.086601} {\bibfield
  {journal} {\bibinfo  {journal} {Phys. Rev. Lett.}\ }\textbf {\bibinfo
  {volume} {100}},\ \bibinfo {pages} {086601} (\bibinfo {year}
  {2008})}\BibitemShut {NoStop}%
\bibitem [{\citenamefont {Dubois}\ \emph
  {et~al.}(2013{\natexlab{a}})\citenamefont {Dubois}, \citenamefont {Jullien},
  \citenamefont {Portier}, \citenamefont {Roche}, \citenamefont {Cavanna},
  \citenamefont {Jin}, \citenamefont {Wegscheider}, \citenamefont {Roulleau},\
  and\ \citenamefont {Glattli}}]{dubois:2013}%
  \BibitemOpen
  \bibfield  {author} {\bibinfo {author} {\bibfnamefont {J.}~\bibnamefont
  {Dubois}}, \bibinfo {author} {\bibfnamefont {T.}~\bibnamefont {Jullien}},
  \bibinfo {author} {\bibfnamefont {F.}~\bibnamefont {Portier}}, \bibinfo
  {author} {\bibfnamefont {P.}~\bibnamefont {Roche}}, \bibinfo {author}
  {\bibfnamefont {A.}~\bibnamefont {Cavanna}}, \bibinfo {author} {\bibfnamefont
  {Y.}~\bibnamefont {Jin}}, \bibinfo {author} {\bibfnamefont {W.}~\bibnamefont
  {Wegscheider}}, \bibinfo {author} {\bibfnamefont {P.}~\bibnamefont
  {Roulleau}}, \ and\ \bibinfo {author} {\bibfnamefont {D.~C.}\ \bibnamefont
  {Glattli}},\ }\href {http://dx.doi.org/10.1038/nature12713} {\bibfield
  {journal} {\bibinfo  {journal} {Nature (London)}\ }\textbf {\bibinfo {volume}
  {502}},\ \bibinfo {pages} {659} (\bibinfo {year}
  {2013}{\natexlab{a}})}\BibitemShut {NoStop}%
\bibitem [{\citenamefont {Levitov}\ \emph {et~al.}(1996)\citenamefont
  {Levitov}, \citenamefont {Lee},\ and\ \citenamefont
  {Lesovik}}]{levitov:1996}%
  \BibitemOpen
  \bibfield  {author} {\bibinfo {author} {\bibfnamefont {L.~S.}\ \bibnamefont
  {Levitov}}, \bibinfo {author} {\bibfnamefont {H.}~\bibnamefont {Lee}}, \ and\
  \bibinfo {author} {\bibfnamefont {G.~B.}\ \bibnamefont {Lesovik}},\ }\href
  {\doibase http://dx.doi.org/10.1063/1.531672} {\bibfield  {journal} {\bibinfo
   {journal} {J. Math. Phys.}\ }\textbf {\bibinfo {volume} {37}},\ \bibinfo
  {pages} {4845} (\bibinfo {year} {1996})}\BibitemShut {NoStop}%
\bibitem [{\citenamefont {Keeling}\ \emph {et~al.}(2006)\citenamefont
  {Keeling}, \citenamefont {Klich},\ and\ \citenamefont
  {Levitov}}]{keeling:2006}%
  \BibitemOpen
  \bibfield  {author} {\bibinfo {author} {\bibfnamefont {J.}~\bibnamefont
  {Keeling}}, \bibinfo {author} {\bibfnamefont {I.}~\bibnamefont {Klich}}, \
  and\ \bibinfo {author} {\bibfnamefont {L.~S.}\ \bibnamefont {Levitov}},\
  }\href {http://link.aps.org/doi/10.1103/PhysRevLett.97.116403} {\bibfield
  {journal} {\bibinfo  {journal} {Phys. Rev. Lett.}\ }\textbf {\bibinfo
  {volume} {97}},\ \bibinfo {pages} {116403} (\bibinfo {year}
  {2006})}\BibitemShut {NoStop}%
\bibitem [{\citenamefont {Haack}\ \emph {et~al.}(2011)\citenamefont {Haack},
  \citenamefont {Moskalets}, \citenamefont {Splettstoesser},\ and\
  \citenamefont {B\"uttiker}}]{haack:2011}%
  \BibitemOpen
  \bibfield  {author} {\bibinfo {author} {\bibfnamefont {G.}~\bibnamefont
  {Haack}}, \bibinfo {author} {\bibfnamefont {M.}~\bibnamefont {Moskalets}},
  \bibinfo {author} {\bibfnamefont {J.}~\bibnamefont {Splettstoesser}}, \ and\
  \bibinfo {author} {\bibfnamefont {M.}~\bibnamefont {B\"uttiker}},\ }\href
  {\doibase 10.1103/PhysRevB.84.081303} {\bibfield  {journal} {\bibinfo
  {journal} {Phys. Rev. B}\ }\textbf {\bibinfo {volume} {84}},\ \bibinfo
  {pages} {081303} (\bibinfo {year} {2011})}\BibitemShut {NoStop}%
\bibitem [{\citenamefont {Haack}\ \emph {et~al.}(2013)\citenamefont {Haack},
  \citenamefont {Moskalets},\ and\ \citenamefont {B\"uttiker}}]{haack:2013}%
  \BibitemOpen
  \bibfield  {author} {\bibinfo {author} {\bibfnamefont {G.}~\bibnamefont
  {Haack}}, \bibinfo {author} {\bibfnamefont {M.}~\bibnamefont {Moskalets}}, \
  and\ \bibinfo {author} {\bibfnamefont {M.}~\bibnamefont {B\"uttiker}},\
  }\href {\doibase 10.1103/PhysRevB.87.201302} {\bibfield  {journal} {\bibinfo
  {journal} {Phys. Rev. B}\ }\textbf {\bibinfo {volume} {87}},\ \bibinfo
  {pages} {201302} (\bibinfo {year} {2013})}\BibitemShut {NoStop}%
\bibitem [{\citenamefont {Ferraro}\ \emph {et~al.}(2013)\citenamefont
  {Ferraro}, \citenamefont {Feller}, \citenamefont {Ghibaudo}, \citenamefont
  {Thibierge}, \citenamefont {Bocquillon}, \citenamefont {F\`eve},
  \citenamefont {Grenier},\ and\ \citenamefont {Degiovanni}}]{ferraro:2013prb}%
  \BibitemOpen
  \bibfield  {author} {\bibinfo {author} {\bibfnamefont {D.}~\bibnamefont
  {Ferraro}}, \bibinfo {author} {\bibfnamefont {A.}~\bibnamefont {Feller}},
  \bibinfo {author} {\bibfnamefont {A.}~\bibnamefont {Ghibaudo}}, \bibinfo
  {author} {\bibfnamefont {E.}~\bibnamefont {Thibierge}}, \bibinfo {author}
  {\bibfnamefont {E.}~\bibnamefont {Bocquillon}}, \bibinfo {author}
  {\bibfnamefont {G.}~\bibnamefont {F\`eve}}, \bibinfo {author} {\bibfnamefont
  {C.}~\bibnamefont {Grenier}}, \ and\ \bibinfo {author} {\bibfnamefont
  {P.}~\bibnamefont {Degiovanni}},\ }\href {\doibase
  10.1103/PhysRevB.88.205303} {\bibfield  {journal} {\bibinfo  {journal} {Phys.
  Rev. B}\ }\textbf {\bibinfo {volume} {88}},\ \bibinfo {pages} {205303}
  (\bibinfo {year} {2013})}\BibitemShut {NoStop}%
\bibitem [{\citenamefont {Juergens}\ \emph {et~al.}(2011)\citenamefont
  {Juergens}, \citenamefont {Splettstoesser},\ and\ \citenamefont
  {Moskalets}}]{juergens:2011}%
  \BibitemOpen
  \bibfield  {author} {\bibinfo {author} {\bibfnamefont {S.}~\bibnamefont
  {Juergens}}, \bibinfo {author} {\bibfnamefont {J.}~\bibnamefont
  {Splettstoesser}}, \ and\ \bibinfo {author} {\bibfnamefont {M.}~\bibnamefont
  {Moskalets}},\ }\href {http://stacks.iop.org/0295-5075/96/i=3/a=37011}
  {\bibfield  {journal} {\bibinfo  {journal} {Europhys. Lett.}\ }\textbf
  {\bibinfo {volume} {96}},\ \bibinfo {pages} {37011} (\bibinfo {year}
  {2011})}\BibitemShut {NoStop}%
\bibitem [{\citenamefont {Rossell\'o}\ \emph {et~al.}(2014)\citenamefont
  {Rossell\'o}, \citenamefont {Battista}, \citenamefont {Moskalets},\ and\
  \citenamefont {Splettstoesser}}]{rossello:2014}%
  \BibitemOpen
  \bibfield  {author} {\bibinfo {author} {\bibfnamefont {G.}~\bibnamefont
  {Rossell\'o}}, \bibinfo {author} {\bibfnamefont {F.}~\bibnamefont
  {Battista}}, \bibinfo {author} {\bibfnamefont {M.}~\bibnamefont {Moskalets}},
  \ and\ \bibinfo {author} {\bibfnamefont {J.}~\bibnamefont {Splettstoesser}},\
  }\href@noop {} {\enquote {\bibinfo {title} {Interference and multi-particle
  effects in a {M}ach-{Z}ehnder interferometer with single-particle sources},}\
  } (\bibinfo {year} {2014}),\ \Eprint {http://arxiv.org/abs/1411.0380}
  {arXiv:1411.0380 [cond-mat]} \BibitemShut {NoStop}%
\bibitem [{\citenamefont {Chung}\ \emph {et~al.}(2007)\citenamefont {Chung},
  \citenamefont {Moskalets},\ and\ \citenamefont {Samuelsson}}]{chung:2007}%
  \BibitemOpen
  \bibfield  {author} {\bibinfo {author} {\bibfnamefont {S.-W.~V.}\
  \bibnamefont {Chung}}, \bibinfo {author} {\bibfnamefont {M.}~\bibnamefont
  {Moskalets}}, \ and\ \bibinfo {author} {\bibfnamefont {P.}~\bibnamefont
  {Samuelsson}},\ }\href {\doibase 10.1103/PhysRevB.75.115332} {\bibfield
  {journal} {\bibinfo  {journal} {Phys. Rev. B}\ }\textbf {\bibinfo {volume}
  {75}},\ \bibinfo {pages} {115332} (\bibinfo {year} {2007})}\BibitemShut
  {NoStop}%
\bibitem [{\citenamefont {Gaury}\ and\ \citenamefont
  {Waintal}(2014)}]{gaury:2014}%
  \BibitemOpen
  \bibfield  {author} {\bibinfo {author} {\bibfnamefont {B.}~\bibnamefont
  {Gaury}}\ and\ \bibinfo {author} {\bibfnamefont {X.}~\bibnamefont
  {Waintal}},\ }\href {http://dx.doi.org/10.1038/ncomms4844} {\bibfield
  {journal} {\bibinfo  {journal} {Nat Commun}\ }\textbf {\bibinfo {volume}
  {5}},\ \bibinfo {pages} {3844} (\bibinfo {year} {2014})}\BibitemShut
  {NoStop}%
\bibitem [{\citenamefont {Chung}\ \emph {et~al.}(2005)\citenamefont {Chung},
  \citenamefont {Samuelsson},\ and\ \citenamefont {B\"uttiker}}]{chung:2005}%
  \BibitemOpen
  \bibfield  {author} {\bibinfo {author} {\bibfnamefont {V.~S.-W.}\
  \bibnamefont {Chung}}, \bibinfo {author} {\bibfnamefont {P.}~\bibnamefont
  {Samuelsson}}, \ and\ \bibinfo {author} {\bibfnamefont {M.}~\bibnamefont
  {B\"uttiker}},\ }\href {\doibase 10.1103/PhysRevB.72.125320} {\bibfield
  {journal} {\bibinfo  {journal} {Phys. Rev. B}\ }\textbf {\bibinfo {volume}
  {72}},\ \bibinfo {pages} {125320} (\bibinfo {year} {2005})}\BibitemShut
  {NoStop}%
\bibitem [{\citenamefont {Battista}\ \emph {et~al.}(2014)\citenamefont
  {Battista}, \citenamefont {Haupt},\ and\ \citenamefont
  {Splettstoesser}}]{battista:2014}%
  \BibitemOpen
  \bibfield  {author} {\bibinfo {author} {\bibfnamefont {F.}~\bibnamefont
  {Battista}}, \bibinfo {author} {\bibfnamefont {F.}~\bibnamefont {Haupt}}, \
  and\ \bibinfo {author} {\bibfnamefont {J.}~\bibnamefont {Splettstoesser}},\
  }\href {\doibase 10.1103/PhysRevB.90.085418} {\bibfield  {journal} {\bibinfo
  {journal} {Phys. Rev. B}\ }\textbf {\bibinfo {volume} {90}},\ \bibinfo
  {pages} {085418} (\bibinfo {year} {2014})}\BibitemShut {NoStop}%
\bibitem [{\citenamefont {Neder}\ \emph {et~al.}(2006)\citenamefont {Neder},
  \citenamefont {Heiblum}, \citenamefont {Levinson}, \citenamefont {Mahalu},\
  and\ \citenamefont {Umansky}}]{neder:2006}%
  \BibitemOpen
  \bibfield  {author} {\bibinfo {author} {\bibfnamefont {I.}~\bibnamefont
  {Neder}}, \bibinfo {author} {\bibfnamefont {M.}~\bibnamefont {Heiblum}},
  \bibinfo {author} {\bibfnamefont {Y.}~\bibnamefont {Levinson}}, \bibinfo
  {author} {\bibfnamefont {D.}~\bibnamefont {Mahalu}}, \ and\ \bibinfo {author}
  {\bibfnamefont {V.}~\bibnamefont {Umansky}},\ }\href {\doibase
  10.1103/PhysRevLett.96.016804} {\bibfield  {journal} {\bibinfo  {journal}
  {Phys. Rev. Lett.}\ }\textbf {\bibinfo {volume} {96}},\ \bibinfo {pages}
  {016804} (\bibinfo {year} {2006})}\BibitemShut {NoStop}%
\bibitem [{\citenamefont {Roulleau}\ \emph {et~al.}(2007)\citenamefont
  {Roulleau}, \citenamefont {Portier}, \citenamefont {Glattli}, \citenamefont
  {Roche}, \citenamefont {Cavanna}, \citenamefont {Faini}, \citenamefont
  {Gennser},\ and\ \citenamefont {Mailly}}]{roulleau:2007}%
  \BibitemOpen
  \bibfield  {author} {\bibinfo {author} {\bibfnamefont {P.}~\bibnamefont
  {Roulleau}}, \bibinfo {author} {\bibfnamefont {F.}~\bibnamefont {Portier}},
  \bibinfo {author} {\bibfnamefont {D.~C.}\ \bibnamefont {Glattli}}, \bibinfo
  {author} {\bibfnamefont {P.}~\bibnamefont {Roche}}, \bibinfo {author}
  {\bibfnamefont {A.}~\bibnamefont {Cavanna}}, \bibinfo {author} {\bibfnamefont
  {G.}~\bibnamefont {Faini}}, \bibinfo {author} {\bibfnamefont
  {U.}~\bibnamefont {Gennser}}, \ and\ \bibinfo {author} {\bibfnamefont
  {D.}~\bibnamefont {Mailly}},\ }\href {\doibase 10.1103/PhysRevB.76.161309}
  {\bibfield  {journal} {\bibinfo  {journal} {Phys. Rev. B}\ }\textbf {\bibinfo
  {volume} {76}},\ \bibinfo {pages} {161309} (\bibinfo {year}
  {2007})}\BibitemShut {NoStop}%
\bibitem [{\citenamefont {Litvin}\ \emph {et~al.}(2008)\citenamefont {Litvin},
  \citenamefont {Helzel}, \citenamefont {Tranitz}, \citenamefont
  {Wegscheider},\ and\ \citenamefont {Strunk}}]{litvin:2008}%
  \BibitemOpen
  \bibfield  {author} {\bibinfo {author} {\bibfnamefont {L.~V.}\ \bibnamefont
  {Litvin}}, \bibinfo {author} {\bibfnamefont {A.}~\bibnamefont {Helzel}},
  \bibinfo {author} {\bibfnamefont {H.-P.}\ \bibnamefont {Tranitz}}, \bibinfo
  {author} {\bibfnamefont {W.}~\bibnamefont {Wegscheider}}, \ and\ \bibinfo
  {author} {\bibfnamefont {C.}~\bibnamefont {Strunk}},\ }\href {\doibase
  10.1103/PhysRevB.78.075303} {\bibfield  {journal} {\bibinfo  {journal} {Phys.
  Rev. B}\ }\textbf {\bibinfo {volume} {78}},\ \bibinfo {pages} {075303}
  (\bibinfo {year} {2008})}\BibitemShut {NoStop}%
\bibitem [{\citenamefont {Kovrizhin}\ and\ \citenamefont
  {Chalker}(2009)}]{kovrizhin:2009}%
  \BibitemOpen
  \bibfield  {author} {\bibinfo {author} {\bibfnamefont {D.~L.}\ \bibnamefont
  {Kovrizhin}}\ and\ \bibinfo {author} {\bibfnamefont {J.~T.}\ \bibnamefont
  {Chalker}},\ }\href {\doibase 10.1103/PhysRevB.80.161306} {\bibfield
  {journal} {\bibinfo  {journal} {Phys. Rev. B}\ }\textbf {\bibinfo {volume}
  {80}},\ \bibinfo {pages} {161306} (\bibinfo {year} {2009})}\BibitemShut
  {NoStop}%
\bibitem [{\citenamefont {Sukhorukov}\ and\ \citenamefont
  {Cheianov}(2007)}]{sukhorukov:2007}%
  \BibitemOpen
  \bibfield  {author} {\bibinfo {author} {\bibfnamefont {E.~V.}\ \bibnamefont
  {Sukhorukov}}\ and\ \bibinfo {author} {\bibfnamefont {V.~V.}\ \bibnamefont
  {Cheianov}},\ }\href {\doibase 10.1103/PhysRevLett.99.156801} {\bibfield
  {journal} {\bibinfo  {journal} {Phys. Rev. Lett.}\ }\textbf {\bibinfo
  {volume} {99}},\ \bibinfo {pages} {156801} (\bibinfo {year}
  {2007})}\BibitemShut {NoStop}%
\bibitem [{\citenamefont {Levkivskyi}\ and\ \citenamefont
  {Sukhorukov}(2008)}]{levkivskyi:2008}%
  \BibitemOpen
  \bibfield  {author} {\bibinfo {author} {\bibfnamefont {I.~P.}\ \bibnamefont
  {Levkivskyi}}\ and\ \bibinfo {author} {\bibfnamefont {E.~V.}\ \bibnamefont
  {Sukhorukov}},\ }\href {\doibase 10.1103/PhysRevB.78.045322} {\bibfield
  {journal} {\bibinfo  {journal} {Phys. Rev. B}\ }\textbf {\bibinfo {volume}
  {78}},\ \bibinfo {pages} {045322} (\bibinfo {year} {2008})}\BibitemShut
  {NoStop}%
\bibitem [{\citenamefont {Moskalets}(2012)}]{moskalets:book}%
  \BibitemOpen
  \bibfield  {author} {\bibinfo {author} {\bibfnamefont {M.~V.}\ \bibnamefont
  {Moskalets}},\ }\href@noop {} {\emph {\bibinfo {title} {Scattering matrix
  approach to non-stationary quantum transport}}}\ (\bibinfo  {publisher}
  {Imperial {C}ollege {P}ress},\ \bibinfo {year} {2012})\BibitemShut {NoStop}%
\bibitem [{\citenamefont {Pilgram}\ \emph {et~al.}(2006)\citenamefont
  {Pilgram}, \citenamefont {Samuelsson}, \citenamefont {F\"orster},\ and\
  \citenamefont {B\"uttiker}}]{pilgram:2006}%
  \BibitemOpen
  \bibfield  {author} {\bibinfo {author} {\bibfnamefont {S.}~\bibnamefont
  {Pilgram}}, \bibinfo {author} {\bibfnamefont {P.}~\bibnamefont {Samuelsson}},
  \bibinfo {author} {\bibfnamefont {H.}~\bibnamefont {F\"orster}}, \ and\
  \bibinfo {author} {\bibfnamefont {M.}~\bibnamefont {B\"uttiker}},\ }\href
  {\doibase 10.1103/PhysRevLett.97.066801} {\bibfield  {journal} {\bibinfo
  {journal} {Phys. Rev. Lett.}\ }\textbf {\bibinfo {volume} {97}},\ \bibinfo
  {pages} {066801} (\bibinfo {year} {2006})}\BibitemShut {NoStop}%
\bibitem [{\citenamefont {F\"orster}\ \emph {et~al.}(2007)\citenamefont
  {F\"orster}, \citenamefont {Samuelsson}, \citenamefont {Pilgram},\ and\
  \citenamefont {B\"uttiker}}]{forster:2007}%
  \BibitemOpen
  \bibfield  {author} {\bibinfo {author} {\bibfnamefont {H.}~\bibnamefont
  {F\"orster}}, \bibinfo {author} {\bibfnamefont {P.}~\bibnamefont
  {Samuelsson}}, \bibinfo {author} {\bibfnamefont {S.}~\bibnamefont {Pilgram}},
  \ and\ \bibinfo {author} {\bibfnamefont {M.}~\bibnamefont {B\"uttiker}},\
  }\href {\doibase 10.1103/PhysRevB.75.035340} {\bibfield  {journal} {\bibinfo
  {journal} {Phys. Rev. B}\ }\textbf {\bibinfo {volume} {75}},\ \bibinfo
  {pages} {035340} (\bibinfo {year} {2007})}\BibitemShut {NoStop}%
\bibitem [{\citenamefont {Marquardt}\ and\ \citenamefont
  {Bruder}(2004{\natexlab{a}})}]{marquardt:2004prl}%
  \BibitemOpen
  \bibfield  {author} {\bibinfo {author} {\bibfnamefont {F.}~\bibnamefont
  {Marquardt}}\ and\ \bibinfo {author} {\bibfnamefont {C.}~\bibnamefont
  {Bruder}},\ }\href {\doibase 10.1103/PhysRevLett.92.056805} {\bibfield
  {journal} {\bibinfo  {journal} {Phys. Rev. Lett.}\ }\textbf {\bibinfo
  {volume} {92}},\ \bibinfo {pages} {056805} (\bibinfo {year}
  {2004}{\natexlab{a}})}\BibitemShut {NoStop}%
\bibitem [{\citenamefont {Marquardt}\ and\ \citenamefont
  {Bruder}(2004{\natexlab{b}})}]{marquardt:2004prb}%
  \BibitemOpen
  \bibfield  {author} {\bibinfo {author} {\bibfnamefont {F.}~\bibnamefont
  {Marquardt}}\ and\ \bibinfo {author} {\bibfnamefont {C.}~\bibnamefont
  {Bruder}},\ }\href {\doibase 10.1103/PhysRevB.70.125305} {\bibfield
  {journal} {\bibinfo  {journal} {Phys. Rev. B}\ }\textbf {\bibinfo {volume}
  {70}},\ \bibinfo {pages} {125305} (\bibinfo {year}
  {2004}{\natexlab{b}})}\BibitemShut {NoStop}%
\bibitem [{\citenamefont {Seelig}\ and\ \citenamefont
  {B\"uttiker}(2001)}]{seelig:2001}%
  \BibitemOpen
  \bibfield  {author} {\bibinfo {author} {\bibfnamefont {G.}~\bibnamefont
  {Seelig}}\ and\ \bibinfo {author} {\bibfnamefont {M.}~\bibnamefont
  {B\"uttiker}},\ }\href {\doibase 10.1103/PhysRevB.64.245313} {\bibfield
  {journal} {\bibinfo  {journal} {Phys. Rev. B}\ }\textbf {\bibinfo {volume}
  {64}},\ \bibinfo {pages} {245313} (\bibinfo {year} {2001})}\BibitemShut
  {NoStop}%
\bibitem [{\citenamefont {F\"orster}\ \emph {et~al.}(2005)\citenamefont
  {F\"orster}, \citenamefont {Pilgram},\ and\ \citenamefont
  {B\"uttiker}}]{forster:2005}%
  \BibitemOpen
  \bibfield  {author} {\bibinfo {author} {\bibfnamefont {H.}~\bibnamefont
  {F\"orster}}, \bibinfo {author} {\bibfnamefont {S.}~\bibnamefont {Pilgram}},
  \ and\ \bibinfo {author} {\bibfnamefont {M.}~\bibnamefont {B\"uttiker}},\
  }\href {\doibase 10.1103/PhysRevB.72.075301} {\bibfield  {journal} {\bibinfo
  {journal} {Phys. Rev. B}\ }\textbf {\bibinfo {volume} {72}},\ \bibinfo
  {pages} {075301} (\bibinfo {year} {2005})}\BibitemShut {NoStop}%
\bibitem [{\citenamefont {Marquardt}(2005)}]{marquardt:2005}%
  \BibitemOpen
  \bibfield  {author} {\bibinfo {author} {\bibfnamefont {F.}~\bibnamefont
  {Marquardt}},\ }\href {http://stacks.iop.org/0295-5075/72/i=5/a=788}
  {\bibfield  {journal} {\bibinfo  {journal} {Europhys. Lett.}\ }\textbf
  {\bibinfo {volume} {72}},\ \bibinfo {pages} {788} (\bibinfo {year}
  {2005})}\BibitemShut {NoStop}%
\bibitem [{\citenamefont {Vanevi\ifmmode~\acute{c}\else \'{c}\fi{}}\ \emph
  {et~al.}(2007)\citenamefont {Vanevi\ifmmode~\acute{c}\else \'{c}\fi{}},
  \citenamefont {Nazarov},\ and\ \citenamefont {Belzig}}]{vanevic:2007}%
  \BibitemOpen
  \bibfield  {author} {\bibinfo {author} {\bibfnamefont {M.}~\bibnamefont
  {Vanevi\ifmmode~\acute{c}\else \'{c}\fi{}}}, \bibinfo {author} {\bibfnamefont
  {Y.~V.}\ \bibnamefont {Nazarov}}, \ and\ \bibinfo {author} {\bibfnamefont
  {W.}~\bibnamefont {Belzig}},\ }\href {\doibase 10.1103/PhysRevLett.99.076601}
  {\bibfield  {journal} {\bibinfo  {journal} {Phys. Rev. Lett.}\ }\textbf
  {\bibinfo {volume} {99}},\ \bibinfo {pages} {076601} (\bibinfo {year}
  {2007})}\BibitemShut {NoStop}%
\bibitem [{\citenamefont {Moskalets}\ and\ \citenamefont
  {B\"uttiker}(2002)}]{moskalets:2002}%
  \BibitemOpen
  \bibfield  {author} {\bibinfo {author} {\bibfnamefont {M.}~\bibnamefont
  {Moskalets}}\ and\ \bibinfo {author} {\bibfnamefont {M.}~\bibnamefont
  {B\"uttiker}},\ }\href {http://link.aps.org/doi/10.1103/PhysRevB.66.205320}
  {\bibfield  {journal} {\bibinfo  {journal} {Phys. Rev. B}\ }\textbf {\bibinfo
  {volume} {66}},\ \bibinfo {pages} {205320} (\bibinfo {year}
  {2002})}\BibitemShut {NoStop}%
\bibitem [{\citenamefont {Pr\^etre}\ \emph {et~al.}(1996)\citenamefont
  {Pr\^etre}, \citenamefont {Thomas},\ and\ \citenamefont
  {B\"uttiker}}]{pretre:1996}%
  \BibitemOpen
  \bibfield  {author} {\bibinfo {author} {\bibfnamefont {A.}~\bibnamefont
  {Pr\^etre}}, \bibinfo {author} {\bibfnamefont {H.}~\bibnamefont {Thomas}}, \
  and\ \bibinfo {author} {\bibfnamefont {M.}~\bibnamefont {B\"uttiker}},\
  }\href {\doibase 10.1103/PhysRevB.54.8130} {\bibfield  {journal} {\bibinfo
  {journal} {Phys. Rev. B}\ }\textbf {\bibinfo {volume} {54}},\ \bibinfo
  {pages} {8130} (\bibinfo {year} {1996})}\BibitemShut {NoStop}%
\bibitem [{\citenamefont {Pedersen}\ and\ \citenamefont
  {B\"uttiker}(1998)}]{pedersen:1998}%
  \BibitemOpen
  \bibfield  {author} {\bibinfo {author} {\bibfnamefont {M.~H.}\ \bibnamefont
  {Pedersen}}\ and\ \bibinfo {author} {\bibfnamefont {M.}~\bibnamefont
  {B\"uttiker}},\ }\href {\doibase 10.1103/PhysRevB.58.12993} {\bibfield
  {journal} {\bibinfo  {journal} {Phys. Rev. B}\ }\textbf {\bibinfo {volume}
  {58}},\ \bibinfo {pages} {12993} (\bibinfo {year} {1998})}\BibitemShut
  {NoStop}%
\bibitem [{\citenamefont {Dubois}\ \emph
  {et~al.}(2013{\natexlab{b}})\citenamefont {Dubois}, \citenamefont {Jullien},
  \citenamefont {Grenier}, \citenamefont {Degiovanni}, \citenamefont
  {Roulleau},\ and\ \citenamefont {Glattli}}]{dubois:2013prb}%
  \BibitemOpen
  \bibfield  {author} {\bibinfo {author} {\bibfnamefont {J.}~\bibnamefont
  {Dubois}}, \bibinfo {author} {\bibfnamefont {T.}~\bibnamefont {Jullien}},
  \bibinfo {author} {\bibfnamefont {C.}~\bibnamefont {Grenier}}, \bibinfo
  {author} {\bibfnamefont {P.}~\bibnamefont {Degiovanni}}, \bibinfo {author}
  {\bibfnamefont {P.}~\bibnamefont {Roulleau}}, \ and\ \bibinfo {author}
  {\bibfnamefont {D.~C.}\ \bibnamefont {Glattli}},\ }\href {\doibase
  10.1103/PhysRevB.88.085301} {\bibfield  {journal} {\bibinfo  {journal} {Phys.
  Rev. B}\ }\textbf {\bibinfo {volume} {88}},\ \bibinfo {pages} {085301}
  (\bibinfo {year} {2013}{\natexlab{b}})}\BibitemShut {NoStop}%
\bibitem [{\citenamefont {Jullien}\ \emph {et~al.}(2014)\citenamefont
  {Jullien}, \citenamefont {Roulleau}, \citenamefont {Roche}, \citenamefont
  {Cavanna}, \citenamefont {Jin},\ and\ \citenamefont
  {Glattli}}]{jullien:2014}%
  \BibitemOpen
  \bibfield  {author} {\bibinfo {author} {\bibfnamefont {T.}~\bibnamefont
  {Jullien}}, \bibinfo {author} {\bibfnamefont {P.}~\bibnamefont {Roulleau}},
  \bibinfo {author} {\bibfnamefont {B.}~\bibnamefont {Roche}}, \bibinfo
  {author} {\bibfnamefont {A.}~\bibnamefont {Cavanna}}, \bibinfo {author}
  {\bibfnamefont {Y.}~\bibnamefont {Jin}}, \ and\ \bibinfo {author}
  {\bibfnamefont {D.~C.}\ \bibnamefont {Glattli}},\ }\href
  {http://dx.doi.org/10.1038/nature13821} {\bibfield  {journal} {\bibinfo
  {journal} {Nature}\ }\textbf {\bibinfo {volume} {514}},\ \bibinfo {pages}
  {603} (\bibinfo {year} {2014})}\BibitemShut {NoStop}%
\bibitem [{\citenamefont {B\"uttiker}(1992)}]{buttiker:1992}%
  \BibitemOpen
  \bibfield  {author} {\bibinfo {author} {\bibfnamefont {M.}~\bibnamefont
  {B\"uttiker}},\ }\href {\doibase 10.1103/PhysRevB.46.12485} {\bibfield
  {journal} {\bibinfo  {journal} {Phys. Rev. B}\ }\textbf {\bibinfo {volume}
  {46}},\ \bibinfo {pages} {12485} (\bibinfo {year} {1992})}\BibitemShut
  {NoStop}%
\bibitem [{\citenamefont {Moskalets}(2014)}]{moskalets:2014}%
  \BibitemOpen
  \bibfield  {author} {\bibinfo {author} {\bibfnamefont {M.}~\bibnamefont
  {Moskalets}},\ }\href {\doibase 10.1103/PhysRevB.90.155453} {\bibfield
  {journal} {\bibinfo  {journal} {Phys. Rev. B}\ }\textbf {\bibinfo {volume}
  {90}},\ \bibinfo {pages} {155453} (\bibinfo {year} {2014})}\BibitemShut
  {NoStop}%
\end{thebibliography}

%


\end{document}